%% file: 00_main.tex
\documentclass[fleqn,usenatbib,useAMS]{mnras}
\usepackage[utf8]{inputenc}
\usepackage{graphicx}
\usepackage{newtxtext,newtxmath}
\usepackage[T1]{fontenc}
\usepackage{float}
\usepackage{hyperref}
\usepackage{amsmath}
\usepackage{amsfonts}
\usepackage{bm}
\usepackage{lscape}
\usepackage[para]{threeparttable}
\usepackage{natbib}
\usepackage{xspace}
\usepackage[normalem]{ulem}
\usepackage[pagewise]{lineno}
\def\medianhostgalmass{10.329\xspace}
\def\samplesize{150\xspace}
\def\grizRMS{0.168\xspace}
\def\griRMS{0.173\xspace}
\def\zRMS{0.195\xspace}
\def\gRMS{0.200\xspace}
\def\rRMS{0.181\xspace}
\def\iRMS{0.175\xspace}

\def\zsigmares{0.157\xspace}
\def\gsigmares{0.145\xspace}
\def\rsigmares{0.152\xspace}
\def\isigmares{0.146\xspace}

\def\grizwRMS{0.159\xspace}
\def\griwRMS{0.164\xspace}
\def\zwRMS{0.183\xspace}
\def\gwRMS{0.186\xspace}
\def\rwRMS{0.175\xspace}
\def\iwRMS{0.165\xspace}

\def\gPeakIntAbsMag{-18.99\xspace}
\def\gPeakIntAbsMagErr{0.03\xspace}
\def\gIntResScat{0.30\xspace}
\def\gIntResScatErr{0.02\xspace}
\def\gExtResScat{0.37\xspace}
\def\gExtResScatErr{0.02\xspace}

\def\rPeakIntAbsMag{-18.92\xspace}
\def\rPeakIntAbsMagErr{0.02\xspace}
\def\rIntResScat{0.26\xspace}
\def\rIntResScatErr{0.02\xspace}
\def\rExtResScat{0.29\xspace}
\def\rExtResScatErr{0.02\xspace}

\def\iPeakIntAbsMag{-18.29\xspace}
\def\iPeakIntAbsMagErr{0.02\xspace}
\def\iIntResScat{0.20\xspace}
\def\iIntResScatErr{0.01\xspace}
\def\iExtResScat{0.22\xspace}
\def\iExtResScatErr{0.01\xspace}

\def\zPeakIntAbsMag{-18.30\xspace}
\def\zPeakIntAbsMagErr{0.02\xspace}
\def\zIntResScat{0.24\xspace}
\def\zIntResScatErr{0.02\xspace}
\def\zExtResScat{0.25\xspace}
\def\zExtResScatErr{0.02\xspace}

\def\gStretchLumSlope{0.174\xspace}
\def\gStretchLumSlopeErr{0.021\xspace}
\def\gStretchLumScat{0.239\xspace}
\def\gStretchLumScatErr{0.015\xspace}

\def\rStretchLumSlope{0.146\xspace}
\def\rStretchLumSlopeErr{0.019\xspace}
\def\rStretchLumScat{0.208\xspace}
\def\rStretchLumScatErr{0.014\xspace}

\def\iStretchLumSlope{0.076\xspace}
\def\iStretchLumSlopeErr{0.017\xspace}
\def\iStretchLumScat{0.190\xspace}
\def\iStretchLumScatErr{0.013\xspace}

\def\zStretchLumSlope{0.143\xspace}
\def\zStretchLumSlopeErr{0.017\xspace}
\def\zStretchLumScat{0.184\xspace}
\def\zStretchLumScatErr{0.013\xspace}

\def\grizMassStepTen{-0.126\xspace}
\def\grizMassStepErrTen{0.027\xspace}

\def\gMassStepTen{-0.099\xspace}
\def\gMassStepErrTen{0.037\xspace}
\def\rMassStepTen{-0.102\xspace}
\def\rMassStepErrTen{0.031\xspace}
\def\iMassStepTen{-0.113\xspace}
\def\iMassStepErrTen{0.028\xspace}
\def\zMassStepTen{-0.105\xspace}
\def\zMassStepErrTen{0.031\xspace}
\def\zSplitMassStepTen{-0.086\xspace}
\def\zSplitMassStepErrTen{0.032\xspace}

\def\grizYSELMScatTen{0.086\xspace}
\def\grizYSELMScatErrTen{0.048\xspace}
\def\grizYSEHMScatTen{0.217\xspace}
\def\grizYSEHMScatErrTen{0.036\xspace}
\def\grizFoundationLMScatTen{0.138\xspace}
\def\grizFoundationLMScatErrTen{0.022\xspace}
\def\grizFoundationHMScatTen{0.103\xspace}
\def\grizFoundationHMScatErrTen{0.013\xspace}

\def\grizMassStepMedian{-0.082\xspace}
\def\grizMassStepErrMedian{0.027\xspace}

\def\gMassStepMedian{-0.065\xspace}
\def\gMassStepErrMedian{0.033\xspace}
\def\rMassStepMedian{-0.065\xspace}
\def\rMassStepErrMedian{0.030\xspace}
\def\iMassStepMedian{-0.080\xspace}
\def\iMassStepErrMedian{0.028\xspace}
\def\zMassStepMedian{-0.074\xspace}
\def\zMassStepErrMedian{0.030\xspace}
\def\zSplitMassStepMedian{-0.052\xspace}
\def\zSplitMassStepErrMedian{0.030\xspace}

\def\NlowTen{49\xspace}
\def\NhighTen{101\xspace}
\def\NMedian{75\xspace}

\title[SNe~Ia in the $z$ band]{Characterising the Standardisation Properties of Type Ia Supernovae in the $z$ band with Hierarchical Bayesian Modelling}

\author[E.~E.~Hayes et al.]{Erin~E.~Hayes,$^{1}$\thanks{e-mail: eeh55@cam.ac.uk}
Suhail~Dhawan,$^{1,2}$
Kaisey~S.~Mandel,$^{1}$
David~O.~Jones,$^{3}$
Ryan~J.~Foley,$^{4}$
Stephen~Thorp,$^{5}$
\newauthor
Matthew~Grayling,$^{1}$
Sam~M.~Ward,$^{1}$
Aaron~Do,$^{1}$
Danial~Langeroodi,$^{6}$
Nicholas~Earl,$^{7}$
Kaylee~M.~de~Soto,$^{8}$
\newauthor
Gautham~Narayan,$^{7,9}$
Katie~Auchettl,$^{4,10}$
Thomas~de~Boer,$^{3}$
Kenneth~C.~Chambers,$^{3}$
David~A.~Coulter,$^{11}$
\newauthor
Christa~Gall,$^{6}$
Hua~Gao,$^{3}$
Luca~Izzo,$^{6}$
Chien-Cheng~Lin,$^{3}$
Eugene~A.~Magnier,$^{3}$
Armin~Rest,$^{11,12}$
\newauthor
and Qinan~Wang$^{13}$
\\
$^{1}$Institute of Astronomy and Kavli Institute for Cosmology, University of Cambridge, Madingley Road, Cambridge CB3 0HA, UK\\
$^{2}$School of Physics and Astronomy, University of Birmingham, Birmingham, UK\\
$^{3}$Institute for Astronomy, University of Hawai'i, 640 N.~A'ohoku Pl., Hilo, HI 96720, USA\\
$^{4}$Department of Astronomy and Astrophysics, University of California, Santa Cruz, CA 95064, USA\\
$^{5}$The Oskar Klein Centre, Department of Physics, Stockholm University, AlbaNova University Centre, SE 106 91 Stockholm, Sweden\\
$^{6}$DARK, Niels Bohr Institute, University of Copenhagen, Jagtvej 128, 2200 Copenhagen, Denmark\\
$^{7}$Department of Astronomy, University of Illinois at Urbana-Champaign, 1002 W. Green St., IL 61801, USA\\
$^{8}$Center for Astrophysics \textbar{} Harvard \& Smithsonian, 60 Garden Street, Cambridge, MA 02138-1516, USA\\
$^{9}$Center for Astrophysical Surveys, National Center for Supercomputing Applications, Urbana, IL 61801, USA\\
$^{10}$School of Physics, The University of Melbourne, VIC 3010, Australia\\
$^{11}$Space Telescope Science Institute, Baltimore, MD 21218, USA\\
$^{12}$Department of Physics and Astronomy, The Johns Hopkins University, Baltimore, MD 21218, USA\\
$^{13}$Department of Physics and Kavli Institute for Astrophysics and Space Research, Massachusetts Institute of Technology, Cambridge, MA 02139, USA}

\date{Accepted 2025 June 26. Received 2025 June 26; in original form 2024 December 6}

\pubyear{2025}

\begin{document}
\label{firstpage}
\pagerange{\pageref{firstpage}--\pageref{lastpage}}
\maketitle

\begin{abstract}
    Type Ia supernovae (SNe~Ia) are \textit{standardisable} candles: their peak magnitudes can be corrected for correlations between light curve properties and their luminosities to precisely estimate distances. Understanding SN~Ia standardisation across wavelength improves methods for correcting SN~Ia magnitudes. Using 150 SNe~Ia from the Foundation Supernova Survey and Young Supernova Experiment, we present the first study focusing on SN~Ia standardisation properties in the $z$ band. Straddling the optical and near-infrared, SN~Ia light in the $z$ band is less sensitive to dust extinction and can be collected alongside the optical on CCDs. Pre-standardisation, SNe~Ia exhibit less residual scatter in $z$-band peak magnitudes than in the $g$ and $r$ bands. SNe~Ia peak $z$-band magnitudes still exhibit a significant dependence on light-curve shape. Post-standardisation, the $z$-band Hubble diagram has a total scatter of RMS $ =\zRMS$ mag. We infer a $z$-band mass step of $\gamma_{z} = \zMassStepTen \pm \zMassStepErrTen$ mag, which is consistent within $1\sigma$ of that estimated from $gri$ data, assuming $R_{V} = 2.61$. When assuming different $R_{V}$ values for high and low mass host galaxies, the $z$-band and optical mass steps remain consistent within $1\sigma$. Based on current statistical precision, these results suggest dust reddening cannot fully explain the mass step. SNe~Ia in the $z$ band exhibit complementary standardisability properties to the optical that can improve distance estimates. Understanding these properties is important for the upcoming Vera Rubin Observatory and \textit{Nancy G.\ Roman Space Telescope}, which will probe the rest-frame $z$ band to redshifts 0.1 and 1.8.
\end{abstract}


\section{Introduction}
\label{sec:intro}
\input{01_introduction}

\section{Data}
\label{sec:data}
\input{02_data}

\section{Methodology}
\label{sec:methods}
\input{03_methods}

\section{Analysis}
\label{sec:analysis}
\input{04_analysis}

\section{Discussion}
\label{sec:discussion}
\input{05_discussion}

\section{Conclusion}
\label{sec:conclusion}
\input{06_conclusion}

\section*{Acknowledgements}
We thank J.\ Johansson for helpful discussions.

Supernova and astrostatistics research at Cambridge University is supported by the European Union’s Horizon 2020 research and innovation programme under European Research Council Grant Agreement No.\ 101002652 (PI K. Mandel) and Marie Skłodowska-Curie Grant Agreement No.\ 873089.

E.E.H.\ is supported by a Gates Cambridge Scholarship (\#OPP1144).
S.D.\ is supported by a Marie Skłodowska Curie Grant Agreement No. 890695, a Kavli Fellowship and a Junior Research Fellowship at Lucy Cavendish College. This work was funded by UK Research and Innovation (UKRI) under the UK government’s Horizon Europe funding Guarantee EP/Z000475/1. 
D.O.J.\ acknowledges support from NSF grants AST-2407632 and AST-2429450, NASA grant 80NSSC24M0023, and HST/JWST grants HST-GO-17128, HST-GO-16269, and JWST-GO-05324, awarded by the Space Telescope Science Institute (STScI), which is operated by the Association of Universities for Research in Astronomy, Inc., for NASA, under contract NAS5-26555.
The UCSC team is supported in part by NASA grants 80NSSC18K0303, 80NSSC18K0303, 80NSSC19K1386, 80NSSC19K0113, 80NSSC20K0953, 80NSSC21K2076, 80NSSC20K0953, 80NSSC22K1513, 80NSSC23K0301, 80NSSC23K1151, and 80NSSC24K1411; NSF grant AST--1815935, the Gordon \& Betty Moore Foundation, the Heising-Simons Foundation, and by a fellowship from the David and Lucile Packard Foundation to R.J.F.
S.T.\ was supported by funding from the European Research Council (ERC) under the European Union's Horizon 2020 research and innovation programmes (grant agreement no. 101018897 CosmicExplorer).
S.M.W.\ was supported by the UK Science and Technology Facilities Council (STFC).
D.L.\ was supported by research grants (VIL16599, VIL54489) from VILLUM FONDEN.
K.M.dS.\ acknowledges support by the NSF through grant AST-2108676 and from LSSTC through grant 2023-SFF-LFI-02-Villar. KMdS thanks the LSST-DA Data Science Fellowship Program, which is funded by LSST-DA, the Brinson Foundation, and the Moore Foundation.
NE acknowledges support from NSF grant AST-2206164.
G.N.\ gratefully acknowledges NSF support from AST-2206195, and CAREER grant AST-2239364, supported in-part by funding from Charles Simonyi, and OAC-2311355, DOE support through the Department of Physics at the University of Illinois, Urbana-Champaign (\#13771275), and support from the HST Guest Observer Program through HST-GO-16764. and HST-GO-17128 (PI: R. Foley).
C.G.\ is supported by research grants (00025501, VIL69896) from VILLUM FONDEN.

The Young Supernova Experiment (YSE) and its research infrastructure is supported by the European Research Council under the European Union's Horizon 2020 research and innovation programme (ERC Grant Agreement 101002652, PI K.\ Mandel), the Heising-Simons Foundation (2018-0913, PI R.\ Foley; 2018-0911, PI R.\ Margutti), NASA (NNG17PX03C, PI R.\ Foley), NSF (AST-1720756, AST-1815935, PI R.\ Foley; AST-1909796, AST-1944985, PI R.\ Margutti), the David \& Lucille Packard Foundation (PI R.\ Foley), VILLUM FONDEN (project 16599, PI J.\ Hjorth), and the Center for AstroPhysical Surveys (CAPS) at the National Center for Supercomputing Applications (NCSA) and the University of Illinois Urbana-Champaign.

Pan-STARRS is a project of the Institute for Astronomy of the University of Hawaii, and is supported by the NASA SSO Near Earth Observation Program under grants 80NSSC18K0971, NNX14AM74G, NNX12AR65G, NNX13AQ47G, NNX08AR22G, 80NSSC21K1572 and by the State of Hawaii. The PS1 Surveys and the PS1 public science archive have been made possible through contributions by the Institute for Astronomy, the University of Hawaii, the Pan-STARRS Project Office, the Max-Planck Society and its participating institutes, the Max Planck Institute for Astronomy, Heidelberg and the Max Planck Institute for Extraterrestrial Physics, Garching, The Johns Hopkins University, Durham University, the University of Edinburgh, the Queen's University Belfast, the Harvard-Smithsonian Center for Astrophysics, the Las Cumbres Observatory Global Telescope Network Incorporated, the National Central University of Taiwan, STScI, NASA under grant NNX08AR22G issued through the Planetary Science Division of the NASA Science Mission Directorate, NSF grant AST-1238877, the University of Maryland, Eotvos Lorand University (ELTE), the Los Alamos National Laboratory, and the Gordon and Betty Moore Foundation.

A major upgrade of the Kast spectrograph on the Shane 3~m telescope at Lick Observatory was made possible through generous gifts from the Heising-Simons Foundation as well as William and Marina Kast. Research at Lick Observatory is partially supported by a generous gift from Google.

The data presented here were obtained (in part) with ALFOSC, which is provided by the Instituto de Astrofisica de Andalucia (IAA) under a joint agreement with the University of Copenhagen and NOT.

YSE-PZ was developed by the UC Santa Cruz Transients Team. The UCSC team is supported in part by NASA grants NNG17PX03C, 80NSSC18K0303, 80NSSC19K0113, 80NSSC19K1386, 80NSSC20K0953, 80NSSC21K2076, 80NSSC22K1513, 80NSSC22K1518, and 80NSSC23K0301; NSF grants AST–1720756, AST–1815935, and AST–1911206; grants associated with Hubble Space Telescope programs DD–14925, DD–15600, GO–15876, GO–16238, SNAP–16239, GO–16690, SNAP–16691, and GO–17128; the Gordon \& Betty Moore Foundation; the Heising-Simons Foundation; fellowships from the Alfred P. Sloan Foundation and the David and Lucile Packard Foundation to RJF.; Gordon and Betty Moore Foundation postdoctoral fellowships and a NASA Einstein fellowship, as administered through the NASA Hubble Fellowship program and grant HST-HF2-51462.001, to D.O.J.; and a National Science Foundation Graduate Research Fellowship, administered through grant No. DGE-1339067, to D.A.C.

\section*{Data Availability}
The code and data needed to reproduce the analysis presented in this work is made publicly available on Github at \url{https://github.com/erinhay/snia-zband/tree/main}. The Foundation Data Release 1 photometry from \citet{Foley_2018} are publicly available from the Monthly Notices of the Royal Astronomical Society at \url{https://doi.org/10.1093/mnras/stx3136}.
The Young Supernova Experiment Data Release 1 from \citet{Aleo_2023} are publicly available on Zenodo at \url{https://doi.org/10.5281/zenodo.7317476}.

\appendix
\section{Data Tables}
\label{sec:appendixA}
\input{07_appendixA.tex}

\section{Sample Selection}
\label{sec:appendixB}
\input{08_appendixB.tex}

\section{Alternative Fitting Methods}
\label{sec:appendixC}
\input{09_appendixC.tex}

\input{tableA1}

\bibliographystyle{mnras}
\bibliography{99_bib}

\label{lastpage}
\end{document}

%% file: 01_introduction.tex
Type Ia supernovae (SNe~Ia) are important cosmological probes: as ``standardizable'' candles, their distances can be precisely inferred from their light curves. Because of this property, SNe~Ia provided the first evidence of an accelerating universe in 1998 \citep{Riess_1998, Perlmutter_1999}. Today, they are being used to place some of the tightest constraints on the expansion rate of the universe, $H_{0}$ \citep{Riess_2022, Freedman_2021}, and the properties of the elusive ``dark energy'' which is causing the accelerated expansion of the universe \citep{DESYR5_2024}. 

With next generation cosmological surveys, such as the Vera C. Rubin Observatory's Legacy Survey of Space and Time \citep[Rubin-LSST;][]{Ivezic_2019} and the Nancy Grace Roman Space Telescope \citep[Roman;][]{Hounsell_2018}, systematic errors will soon dominate over statistics in the cosmological parameter inference error budget. The majority of the systematic uncertainties in the estimates of cosmological distance moduli from SNe~Ia come from the standardisation of SN~Ia light curves. One approach to improving models of SNe~Ia light curves for cosmology is by studying SNe~Ia light curves as a function of wavelength. Firstly, understanding how standard SNe~Ia are in different filters may reveal that SNe~Ia in certain wavelength regimes are relatively better standard candles (i.e. exhibit a narrower distribution of peak absolute magnitudes). Characterising these properties can inform how to dedicate observational follow-up of discovered events. Secondly, the properties of SNe~Ia across wavelength can improve our knowledge of the astrophysics underlying these systems, such as properties of the dust in the host galaxy of a SN~Ia. In this work, we present the first comprehensive analysis of the standardisation properties of SNe~Ia in the $z$ band.

The process of standardising of SN~Ia light curves involves correcting for a number of well-documented correlations between SN~Ia luminosities and light curve shapes, colours, and host galaxy properties. These correlations include the broader-brighter relation \citep{Phillips_1993, Guy_2007, Mandel_2011, Burns_2014}, which accounts for the more luminous nature of SNe~Ia which have slower declining light curves, and effective apparent colour-luminosity relation \citep{Tripp_1998}, which can be decomposed \citep{Mandel_2017} into a correction for the intrinsically redder colour of less luminous SNe~Ia and interstellar reddening due to dust in the host galaxy \citep{Burns_2014, Burns_2018, Thorp_2021, Mandel_2022}. Many cosmological analyses also correct for the so-called ``mass step'' -- the tendency of SNe~Ia in high-mass galaxies to be systematically brighter post-standardisation than those in low-mass galaxies \citep{Kelly_2010, Sullivan_2010}. Although the mass step is well-documented in the literature, its physical origin remains an open question.

To date, the samples of SNe~Ia used in cosmological analyses have mainly, though not exclusively, been from optical telescopes \citep{Jones_2018_PScosmo, Riess_2022, Rubin_2023, DESYR5_2024}. Because of the variation observed in SNe~Ia in the optical, standardisation corrections are necessary for precise cosmological analyses. However, SNe~Ia observed in the near-infrared (NIR) wavelength regime have been suggested to be relatively better standard candles compared to optical wavelengths at peak \textit{pre-standardisation}, i.e. prior to corrections for known light curve decline-rate and colour correlations with luminosity \citep{Krisciunas_2004, Nobili_2005, Wood_Vasey_2008, Mandel_2009, Folatelli_2010, Kattner_2012, BaroneNugent_2012, Weyant_2014, Stanishev_2018, Burns_2018, Dhawan_2018, Avelino_2019, Jones_2022, Peterson_2024}. Specifically, the peak NIR magnitudes of SNe~Ia pre-standardisation tend to show less residual scatter and a weaker dependence on the decline-rate parameter than in the optical. These studies may be sensitive to the sample and model used in the analysis, though \citep{Do_2024}.

Light in the NIR also benefits from being 5 to 11 times less sensitive to dust extinction than the optical \citep[e.g.][]{Fitzpatrick_1999}. Because of the difficulty of disentangling the extrinsic host galaxy dust effects from variations in the intrinsic colour populations of SNe~Ia \citep{Thorp_2021, Mandel_2022, TM_2022}, dust in the host galaxy is a significant source of uncertainty in the standardisation of SNe~Ia \citep{Freedman_2009}. Therefore, if indeed SNe~Ia exhibit a smaller intrinsic dispersion in their NIR peak magnitudes, this property and the increased transparency of the NIR to dust would make distance moduli estimates to SNe~Ia from NIR observations more robust to these systematic uncertainties.

The weaker sensitivity to dust also makes the NIR well-suited to study the mass step, which has been suggested to arise from different dust law shapes, parameterised by $R_{V}$, in high and low mass host galaxies \citep{Brout_Scolnic_2021, Popovic_2021_dust}. If dust is the dominant driver of the mass step, a reduced mass step is expected in the Hubble residuals from distance estimates based on NIR data because of the reduced sensitivity of this wavelength regime to dust.\footnote{We note that a reduced mass step in the NIR relative to the optical does not definitively point to dust as the dominant driver of the mass step. It may be possible for intrinsic differences between progenitor systems of SNe~Ia in high and low mass host galaxies \citep{Childress_2013_progenitors} to lead to a reduced mass step in the NIR.} Indeed, \citet{Peterson_2024} recently used simulations to examine the expected size of the mass step under the assumption of the dust explanation first put forward by \citet{Brout_Scolnic_2021} and found a reduced, though still non-zero, mass step in the NIR. Therefore, there has recently been elevated interest in studying SN~Ia in the NIR to probe the origin of the mass step \citep{Uddin_2020, Ponder_2021, Johansson_2021, Thorp_2021, Jones_2022, TM_2022, Uddin_2023, Grayling_2024, Peterson_2024, Thorp_2024}.

Several works have examined the contribution of dust to the inferred mass step by estimating the mass step in individual bands from the optical to the NIR. The results from these analyses have varied from detections of non-zero mass steps that are consistent with the optical mass step in one or more NIR bands \citep{Uddin_2020, Ponder_2021, Jones_2022, Uddin_2023} to a mass step in the NIR that is consistent with both zero and with the optical mass step \citep{Johansson_2021, TM_2022, Peterson_2024}. The lack of evidence for a decrease in the size of the NIR mass step relative to the optical broadly suggests that the mass step cannot be explained in full by dust \citep[see e.g.,][for alternative explanations of the mass step]{Childress_2013_progenitors}. However, the hypothesis that differing properties of dust in high and low mass host galaxies explains the mass step cannot be ruled out due to the limited statistical precision of these studies. 

The $z$ band straddles the optical and NIR wavelength regimes (rest-frame $\sim8000-9500$\AA). Light in the $z$ band is 2.6x less sensitive to dust than optical bands. Yet, the standardisation properties and mass step of SNe~Ia in the $z$ band have not been explored in the literature. If SNe~Ia prove to be relatively better standard candles in the $z$ band, the advantage of this wavelength regime would be compounded by the fact that $z$-band data is more easily obtainable than data in the NIR. In particular, $z$-band data can be collected alongside the optical using charge coupled devices (CCDs). The brightness of the atmosphere in the NIR makes it difficult to obtain NIR observations from the ground to a comparable depth to the optical with CCDs. Therefore, NIR observations of SNe~Ia must be collected as follow-up of SNe~Ia that have already been discovered, subjecting these samples of SNe~Ia to complex selection effects as a result. The exact consequences of these selection effects on cosmological inference are not well understood, and \citet{Rigault_2020} cautions that the unknown impact of selection effects may already be biasing our inference of cosmological parameters. 

Given these strengths of the $z$ band, we set out to determine the standardisation properties of SNe~Ia in the $z$ band. For this study, we use a spectroscopically confirmed sample of SN~Ia observed in the $griz$ filters from the Foundation Supernova Survey Data Release 1 \citep{Foley_2018} and the Young Supernova Experiment Data Release 1 \citep{Aleo_2023}. We fit the $griz$ light curves from this sample with \textsc{BayeSN}, a hierarchical Bayesian model for the spectral energy distributions (SEDs) of SNe~Ia in the $gBVrizYJH$ filters \citep{Mandel_2022, Grayling_2024}. This work represents the first investigation of the standardisation properties of SNe~Ia in an individual band using \textsc{BayeSN}, including an individual-band mass step estimate, and the first $z$-band Hubble diagram.

Combining optical data with data in redder bands leads to the tightest constraints on distance estimates for SNe~Ia \citep{Mandel_2022, Dhawan_2023}. Yet, the potential of the $z$ band to increase the precision of SNe~Ia standardisation has not been explored in the literature to date. Improving our physical understanding of the empirical corrections applied to SN~Ia light curves is of the utmost importance, as demonstrated by the current $5\sigma$ tension between early- and late-time estimates of $H_{0}$ \citep{Planck_2020, Riess_2022}. This tension puts pressure on $\Lambda$CDM and, if determined to be physical, may be suggestive of new physics \citep[e.g.,][for a review]{Mortsell_2018, DiValentino_2021}. Furthermore, measurements of the present-day dark energy equation of state, $w_{0}$, and its evolution with time, $w_{a}$ from \citet{DESI_2024, DESYR5_2024, DESI_2025} recently suggested deviations from $\Lambda$CDM. To sufficiently rule out the possibility that these tensions arise from an empirical error \citep{Dhawan_2025}, uncertainties in SN~Ia estimates of cosmological parameters must be reduced from the current 1.5\% level to percent level precision \citep{Riess_2022, DESYR5_2024}.

The paper is structured as follows. $\S$\ref{sec:data} describes the data and sample selection cuts. In $\S$\ref{sec:methods}, we describe the methodology for fitting light curves and modelling the shape-luminosity relation and mass step. We present our findings in $\S$\ref{sec:analysis} and discuss these results in $\S$\ref{sec:discussion}. In $\S$\ref{sec:conclusion}, we conclude. Throughout this analysis, we assume a fiducial flat cosmology with $H_{0} = 73.24$ and $\Omega_m = 0.28$, as was adopted for the low-z training of \textsc{BayeSN} \citep{Mandel_2022, Grayling_2024}. The code used in this work can be found on GitHub at \url{https://github.com/erinhay/snia-zband}.

%% file: 02_data.tex
Data for this project comes from two surveys which were both carried out on the Panoramic Survey Telescope and Rapid Response System (Pan-STARRS) telescope \citep{Chambers_2016}. The Foundation Supernova Survey (Foundation), which ran from 2015-2017, produced a sample of 225 SN~Ia \citep{Foley_2018}. The first data release from the ongoing Young Supernova Experiment (YSE), which had its first observations in 2019, produced an additional 292 spectroscopically confirmed SN~Ia \citep{Aleo_2023}.

As this work is the first Hubble diagram analysis from YSE DR1, we take additional steps to clean the YSE DR1 data to prepare it for cosmology. In particular, we remove observations that have been flagged as having masked central pixels, as a mask through the central pixel severely impacts the flux measurement for that observation. We also require that the measurement uncertainty for each observation be less than 0.3 mag. These cuts remove 7.93\% of the total number of observations within the range of (-10, 40) phase days that is fit by \textsc{BayeSN}.

In addition, we update the redshifts and redshift uncertainties from the estimates reported in YSE DR1. In YSE DR1, a conservative redshift uncertainty of 0.01 was assumed for all spectroscopically-classified objects, though the redshift measurements themselves as reported on the NASA/IPAC Extragalactic Database\footnote{\url{https://ned.ipac.caltech.edu/}} and Transient Name Server\footnote{\url{https://www.wis-tns.org/}} were often of higher precision. If the SN~Ia is matched to a host galaxy with a spectroscopic redshift, we proceed with the redshift and redshift uncertainty from the host. For the remaining 150 SNe~Ia, we use the redshift estimate and uncertainty from the transient's spectra. For the 66 SNe~Ia that do not have reported redshift measurement uncertainties, we assume a conservative uncertainty of 0.01. The final versions of the data used in this analysis can be found on the GitHub repository associated with this paper. In Appendix \ref{sec:appendixA}, we provide an abridged table that includes the information necessary to reproduce the analysis in this paper. The full version of this table is provided online.

\subsection{Host Galaxy Mass Estimates}
The host galaxy masses are estimated using \texttt{Blast} \citep{Jones_2024}. \texttt{Blast} is a web application that matches a specified transient to its host galaxy, retrieves all available archival data for the host, and uses that data to estimate host properties via SED fitting. The archival host photometry that is available for the transients comes from one or some of the Two Micron All-Sky Survey \citep[2MASS;][]{Skrutskie_2006}, the Wide-field Infrared Survey Explorer \citep[WISE;][]{Wright_2010}, the Galaxy Evolution Explorer \citep[GALEX;][]{Martin_2005}, Pan-STARRS \citep{Chambers_2016, Flewelling_2020, Magnier_2020}, SDSS \citep{York_2000, Blanton_2017, Ahumada_2020}, and/or the DESI Legacy Imaging Surveys \citep{Dey_2019}. The photometry is fit using \texttt{prospector}, a tool for estimating stellar population properties via SED fitting with a flexible suite of models \citep{Johnson_2021}. \texttt{Blast} uses the Prospector-$\alpha$ model within \texttt{prospector}, which uses a non-parametric star-formation history to model the host galaxy SED \citep{Leja_2017}. Priors on all parameters are as listed in \citet{Jones_2024}. The posterior is sampled using simulation-based inference (SBI) via the SBI$++$ neural posterior estimator \citep{Wang2023}, which is trained on large simulated realizations of low-$z$ galaxies ($0 < z < 0.2$) and uses normalizing flows to efficiently sample the posterior distribution. The \texttt{prospector} estimates of the mass of the host galaxies for the SNe~Ia are then made available through \texttt{Blast}\footnote{\url{https://blast.scimma.org/}.}.

\input{table1}

\subsection{Sample Selection}
\label{sec:data-cuts}
We define our cosmological sample using cuts on spectroscopic classification, data quality, and \textsc{BayeSN} parameter estimates. We first remove the peculiar SNe~Ia from our sample, which includes SNe~Ia-91bg-like, SNe~Ia-91T-like, SNe~Ia-CSM, and SNe~Iax. There is ongoing research into the cosmological utility of SNe~Ia-91bg-like and SNe~Ia-91T-like, which both have peculiar spectra relative to ``normal'' SNe~Ia. Though some analyses have suggested that SNe~Ia-91bg-like may be standardisable with an appropriate choice of the shape parameterisation \citep{Burns_2018, Graur_2024}, we follow the precedent of past cosmological analyses \citep[e.g.][]{Rest_2014, Betoule_2014, Foley_2018} and remove the SNe~Ia-91bg-like from our sample. While SNe~Ia-91T-like have been used in cosmological analyses, recent work has suggested this sub-class of SNe~Ia are over-luminous post-standardisation \citep{Phillips_2022}, so we also remove these objects from the sample.

Of the remaining 446 Foundation and YSE objects, we remove 102 additional SNe~Ia outside the redshift range of $0.015 < z < 0.08$ for Foundation and $0.015 < z < 0.1$ for YSE to minimise Malmquist bias. The redshift limits for Foundation are adopted from previous cosmological analyses of this sample \citep{Foley_2018, Jones_2019, Thorp_2021}. To determine the redshift limits for YSE, we apply a KS test to the distributions of inferred light curve parameters -- $\theta$ and $A_{V}$ (see \S\ref{sec:bayesn} for more details on the \textsc{BayeSN} model parameters) -- in redshift bins of $dz = 0.01$ relative to reference distributions from the SNe~Ia in the range $0.015 < z < 0.05$. The distributions of $\theta$ and $A_{V}$ within each bin remain consistent with being drawn from the same distribution as the reference distribution up to redshift $z=0.1$. Above this redshift, the distributions of these inferred parameters are found to be inconsistent with the reference distribution, indicating a change in population that likely originates from Malmquist bias. We find that our results are insensitive to the choice of redshift bin size.

We next impose a requirement on the quality of the light curve data. We require at least one observation in each of the optical $gri$ bands. We also require that in at least one of the optical bands, there be an observation before the time of maximum flux in the $B$ band, $T_{0}$\footnote{The \textsc{BayeSN} model is defined such that $T_{0}$ represents the time of maximum flux in the Bessel $B$ band, similarly to other SN~Ia SED and light curve models \citep{Hsiao2007, Guy_2007, Taylor_2021, Kenworthy_2021, Pierel_2022, Burns_2018}. This choice anchors the template in time, but does not affect the inference of other model parameters.}. Regarding the $z$-band data, we require there be 3 observations in the $z$ band within (-10, 40) phase days to ensure we can get a reliable estimate of the distance modulus from a \textsc{BayeSN} fit to the $z$-band only light curve. For further details on the $z$-band data quality cut, see Appendix \ref{sec:appendixA}. To ensure high quality light curve fits, we include only those objects with a reduced $\chi^{2}_{griz} < 3$\footnote{\textsc{BayeSN} provides a posterior on the model parameters, rather than a maximum likelihood estimate from a parameter optimisation scheme. To retrieve the model magnitudes to estimate $\chi^{2}$, we sample from the posterior 100 times, simulate light curves for each draw of the parameters, and take the mean of the simulated light curves at the times of observation. The \textsc{BayeSN} fit to the $griz$ data, as described in \S\ref{sec:bayesn}, is used to estimate $\chi^{2}$.}. Finally, we require uncertainties on $T_{0}$ of less than a day for the \textsc{BayeSN} fit to the full $griz$ light curve data, as well as the fit to the optical only $gri$ data. We also enforce that the estimates of $T_{0}$ agree across these two fits to within 0.5 days.

To ensure the sample lies within the bounds of the original training set of the model, we impose cuts on the \textsc{BayeSN} light curve parameters. In particular, we make cuts on the light curve shape parameter, $-1.5 < \theta < 3$, and the host dust extinction, $A_{V} < 1$. We describe the \textsc{BayeSN} model in more detail in \S\ref{sec:bayesn}. The final sample is left at \samplesize objects, with 41 from YSE and 105 from Foundation. The full set of cuts is summarised in Table \ref{tab:datacuts}.

\begin{figure}
    \centering
    \includegraphics[width=0.95\linewidth]{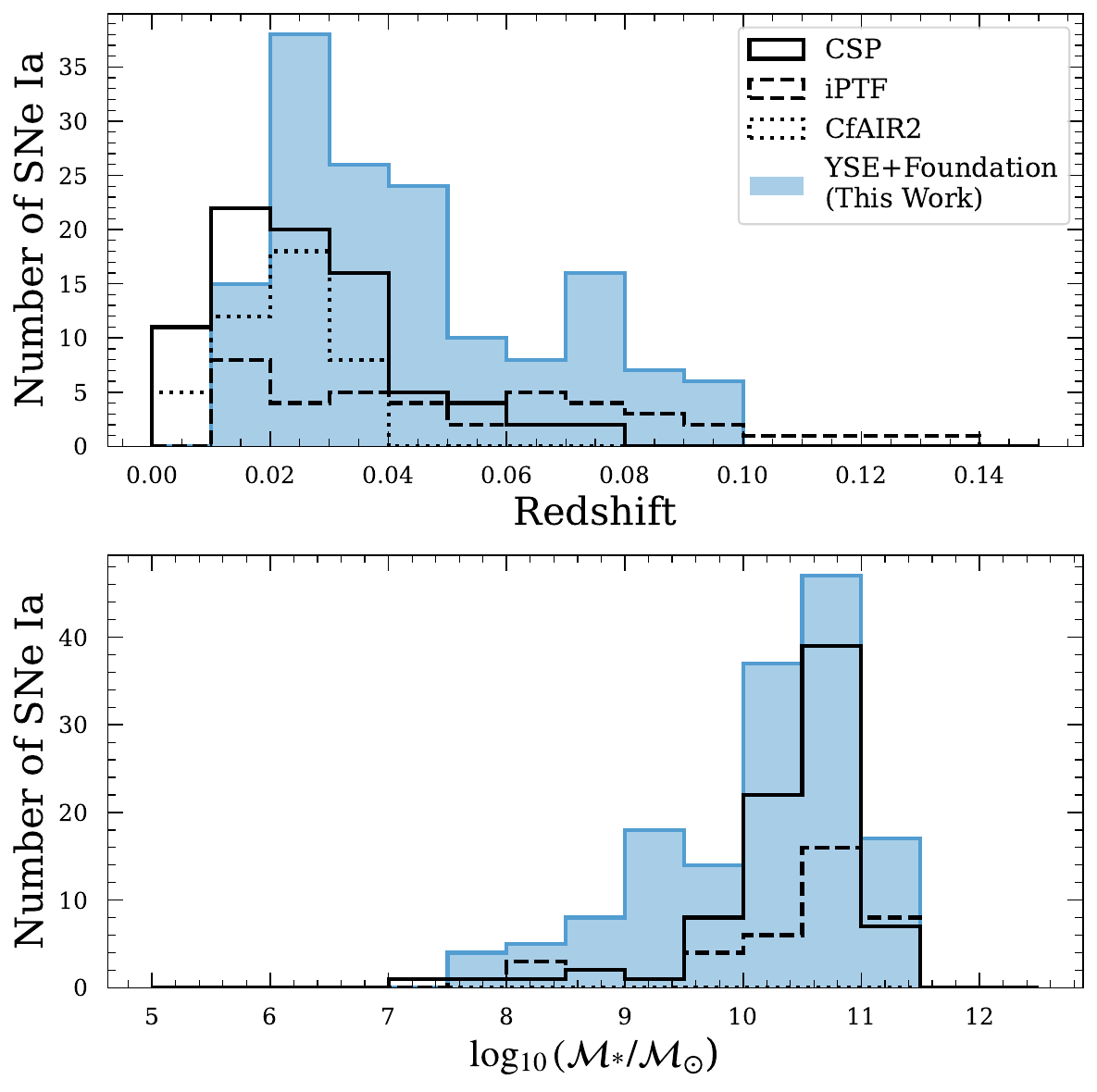}
    \caption{The distribution of redshifts (top) and host galaxy masses (bottom) for the combined YSE and Foundation sample (shaded blue) compared to that for the CSP-I \citep[solid black line;][]{Contreras_2010, Stritzinger_2011}, iPTF \citep[dashed black line;][]{Johansson_2021}, and CfAIR2 \citep[dotted black line;][]{Wood_Vasey_2008, Friedman_2015} NIR samples.}
    \label{fig:pop-dists}
\end{figure}

\begin{figure}
    \centering
    \includegraphics[width=\linewidth]{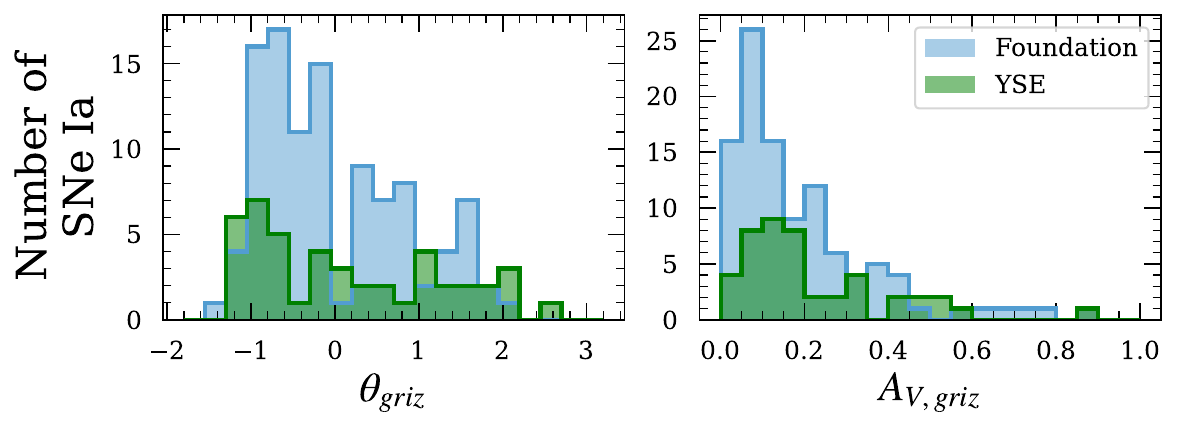}
    \caption{The distributions of $\theta$ (left) and $A_{V}$ (right) for the sample of \samplesize SNe~Ia by survey. The estimates of $\theta$ and $A_{V}$ for each object come from the \textsc{BayeSN} fit to the $griz$ light curve data, as discussed in \S\ref{sec:methods}.}
    \label{fig:theta-av-dists}
\end{figure}

\begin{figure}
    \centering
    \includegraphics[width=0.95\linewidth]{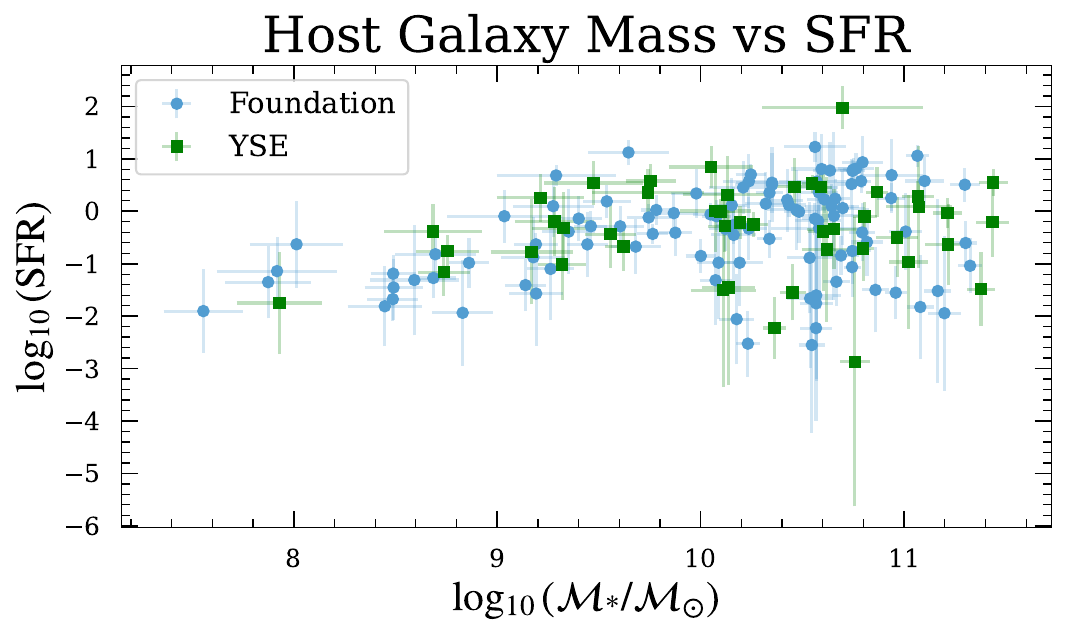}
    \caption{The star formation rate (SFR) vs.\ stellar mass for the host galaxies of the final sample of YSE and Foundation SNe~Ia. The markers show the mean and the error bars show the standard deviation of the posterior samples from the Prospector-$\alpha$ fit \citep{Leja_2017, Johnson_2021} from \texttt{Blast} \citep{Jones_2024}.}
    \label{fig:mass-sfr}
\end{figure}

The distribution of redshifts and of host galaxy masses in our sample is shown in Figure \ref{fig:pop-dists}. In the final sample, 132 SNe~Ia have redshifts from a host galaxy spectra and 17 have redshifts from a SN spectra. Of those 17 SNe~Ia, 14 have redshift measurement uncertainties of 0.01 and 3 have redshift measurement uncertainties of 0.001. In this figure, we additionally compare the sample from this work to NIR samples used in similar studies from the Harvard-Smithsonian Center for Astrophysics \citep[CfAIR2;][]{Wood_Vasey_2008, Friedman_2015}, the intermediate Palomar Transient Facility \citep[iPTF;][]{Johansson_2021}, and the Carnegie Supernova Project I \citep[CSP-I;][]{Contreras_2010, Stritzinger_2011, Krisciunas_2007}. In Figure \ref{fig:theta-av-dists}, we show the distribution of $\theta$ and $A_{V}$ for the final sample of \samplesize objects and in Figure \ref{fig:mass-sfr}, we show the distribution of stellar mass vs. star formation rate (SFR) for the host galaxies of the final sample of SNe~Ia. 

%% file: table1.tex
\begin{table}
    \centering
    \caption{The number of objects cut and remaining in the sample following each data cut imposed on the data set to ensure a pure cosmological sample of SN~Ia. The final sample consists of \samplesize objects.}
    \begin{threeparttable}
        \begin{tabular}{c|c|c|c|c}
            & \multicolumn{2}{c}{\textbf{YSE}} & \multicolumn{2}{c}{\textbf{Foundation}} \\
            \hline
             & $N_{\text{cut}}$ & $N_{\text{rem}}$ & $N_{\text{cut}}$ & $N_{\text{rem}}$ \\
            \hline \hline
            \textbf{Total Spec Sample} & -- & 292 & -- & 225 \\
            Normal Ias & 26 & 266 & 45 & 180 \\
            Redshift Cut\tnote{a} & 81 & 185 & 18 & 162 \\
            $gri$ Data Requirement\tnote{b} & 66 & 119 & 47 & 115 \\
            $z$-band Data Requirement\tnote{c} & 62 & 57 & 1 & 114 \\
            Reduced $\chi^{2} < 3$ & 3 & 54 & 3 & 111 \\
            $T_{0}$ Cut\tnote{d} & 8 & 46 & 6 & 105 \\
            $A_{V} < 1$ & 0 & 46 & 0 & 105 \\
            $-1.5 < \theta < 3$ & 1 & 45 & 0 & 105 \\
            \hline
            \textbf{Final Sample} & & 45 & & 105 \\
            \hline
            \hline
        \end{tabular}
        \begin{tablenotes}\footnotesize
            \item[a] The ``redshift cut'' requires that YSE SNe~Ia have $0.015 < z < 0.1$ and the Foundation SNe~Ia have $0.015 < z < 0.08$.
            
            \item[b] The ``$gri$ data requirement'' requires that there be at least one observations in each of the $gri$ bands and at least one observation in any of the $gri$ bands before peak.

            \item[c] The ``$z$-band data requirement'' requires that there be 3 observations in the $z$ band.

            \item[d] The ``$T_{0}$ Cut'' requires that the uncertainty on $T_{0}$ be less than a day for both the $griz$ and $gri$ fits and that $T_{0}$ from the $griz$  and $gri$ fits agree within 0.5 days.
        \end{tablenotes}
    \end{threeparttable}
    \label{tab:datacuts}
\end{table}

%% file: 03_methods.tex
\subsection{Light Curve Fitting with \textsc{BayeSN}}
\label{sec:bayesn}
\textsc{BayeSN} is a hierarchical Bayesian model for SN~Ia spectral energy distributions (SEDs) in the optical to infrared from 3500 \r{A} to 18000 \r{A} \citep{Thorp_2021, Mandel_2022, Grayling_2024}. \textsc{BayeSN} provides robust and continuous fits of SN~Ia light curves with physically distinct parameters for intrinsic properties of the SN~Ia and host galaxy dust effects. These parameters include a light curve shape parameter, $\theta$, which encodes the broader-brighter relationship. Further intrinsic colour variations among SNe~Ia are captured by a set of fitted time- and wavelength-varying components, $\epsilon(\lambda, t)$. The amount of reddening and extinction from host galaxy dust are quantified by $A_{V}$, the host dust extinction, and $R_{V}$, the dust law shape parameter, according to the \citet{Fitzpatrick_1999} dust law. Accounting for these effects, \textsc{BayeSN} estimates a distance modulus, $\mu$, which has been corrected for shape and colour effects, from the SN~Ia photometry. The \textsc{BayeSN} model and priors on the model parameters are described in more detail in \citet{Mandel_2022}.

For this analysis, we fit with the ``T21'' \textsc{BayeSN} model, trained within the range of 3500-9500 \r{A}, described in \citet{Thorp_2021}. This model was trained on Foundation data in the $griz$ filters, so it is ideal for use on this data set, which is composed predominantly of Foundation data. Unless a different host galaxy dust treatment is specified, we assume a fixed $R_{V} = 2.61$ for all SNe~Ia -- the population mean $R_{V}$ learned in the training of the T21 model. We fit for residual scatter within our hierarchical Bayesian models, described in \S\ref{sec:sampling}, for the population distributions of peak magnitudes, the shape-luminosity relation, and the mass step after fitting with \textsc{BayeSN}, so we do not include the default achromatic scatter variance term, $\sigma_{0}^{2}$, in the T21 model when fitting the light curves, instead setting it to zero.

We use the \textsc{BayeSN} fits to the data in two ways: 1) to estimate the peak apparent magnitudes in an individual filter\footnote{We can use \textsc{BayeSN} to recover the inferred peak apparent magnitude of a SN~Ia $s$ in some band. To do so, we take a sample from the posterior and simulate a rest-frame light curve from \textsc{BayeSN} with those parameter values, then extract the rest-frame apparent magnitude in the band of interest at $T_{0}$, the time of maximum $B$-band flux. The mean and uncertainty of the peak magnitude is given by the mean and standard deviation for 500 posterior samples. The peak apparent magnitudes as obtained in this way are not corrected for shape, colour, or dust effects. Rather, when used in this way, \textsc{BayeSN} can be thought of as interpolating the SN~Ia light curve, without adjusting the resulting peak apparent magnitude to account for standardisation corrections. We add uncertainty arising from redshift-based cosmological distances in quadrature to the uncertainty on the peak magnitude.}, and 2) to estimate the  ``corrected'' distance modulus from the SNe~Ia light curves. Here, we use ``corrected'' to denote that both intrinsic (i.e. shape and colour relations) and extrinsic (i.e. dust) corrections, as learned in the training of the model, have been applied to give the most precise distance modulus estimate for each SN~Ia.

\begin{figure}
    \centering
    \includegraphics[width=\linewidth]{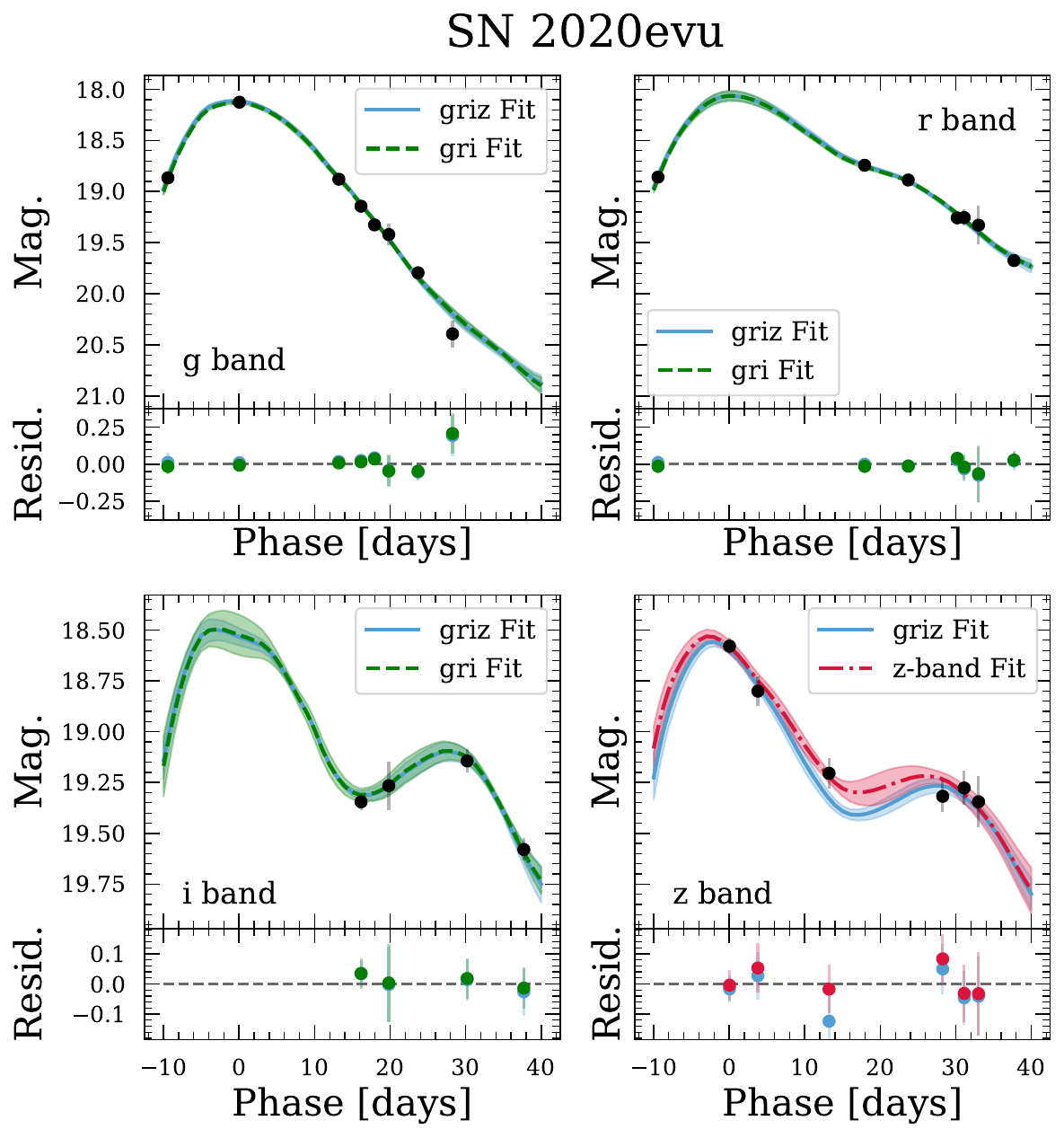}
    \caption{An SN~Ia at $z=0.059$ observed by YSE. Each subplot shows the data in an individual filter (upper right: $g$ band, upper left: $r$ band, lower right: $i$ band, lower left: $z$ band). The \textsc{BayeSN} fit to the $griz$ data is shown with a blue solid line, the fit to the $gri$ data is shown with a green dashed line, and the fit to the $z$-band only data is shown with an orange dot-dashed line. The YSE light curves that pass the data cuts have a median of 5 $z$-band observations. The Foundation light curves that pass the data cuts have a median of 6 $z$-band observations.}
    \label{fig:lc}
\end{figure}

We carry out six different \textsc{BayeSN} fits of each light curve:
\begin{itemize}
    \item individual-band: the fit to an individual-band light curve with all intrinsic parameters ($\theta$, $\epsilon$) being fit for, $A_{V}$ fixed to zero, and $T_{0}$ fixed to the value from an initial fit to the $gri$ data.

    \item $griz$: a fit to the full $griz$ light curve, with all parameters being fit for, and with $T_{0}$ fixed to the value from an initial fit to the $griz$ data,

    \item $gri$: a fit to the optical $gri$ only data, with all parameters being fit for, and with $T_{0}$ fixed to the value from an initial fit to the $gri$ data,
\end{itemize}
For the $griz$ and $gri$ fits, we opt to use $T_{0}$ as estimated from the $griz$ and $gri$ data, respectively, to maintain consistency in the data that is used in these fits. For the fit to the individual-band data, we use $T_{0}$ from an initial fit to the $gri$ data. This choice maintains the use of optical only data for the individual optical filter fits. For the $z$-band fit, $T_{0}$ is anchored with completely independent $gri$ data. An example light curve with the $griz$, $gri$, and $z$-band only fits is shown in Figure \ref{fig:lc}.

The individual-band fits as described do not include a correction for host galaxy dust extinction. In an individual band, there is no colour information to constrain the effects of dust extinction. Without multi-band information, then, the posterior on $A_{V}$ will revert to the prior, which may not constrain well the dust effects for an individual SN~Ia. For this reason, we fix $A_{V}$ to zero in the individual-band \textsc{BayeSN} fits to obtain dust-extinguished estimates. Additionally, to compute the individual-band magnitudes corrected for host galaxy extinction, we use $A_{V}$ from the \textsc{BayeSN} fit to the $griz$ data for each SN~Ia. We assume dust extinction follows a \citet{Fitzpatrick_1999} dust law evaluated at the effective wavelength of the relevant Pan-STARRS band as implemented in the \texttt{extinction} package \citep{Barbary_2017}. By default, we use $R_{V} = 2.61$, which was determined in training of the T21 model \citep{Thorp_2021}, although in \S\ref{sec:analysis-massstep}, we also explore the impact of host mass-dependent $R_{V}$ values on our mass step inferences. 

There are alternative choices that can be made when fitting individual-band data with \textsc{BayeSN}, such as fixing $\theta$. To ensure that the model best fits the data in an individual band, we proceed with a fit which only uses data from that band. We test the effects of alternative fitting choices on the analysis in Appendix \ref{sec:appendixB} for completeness.

\subsection{Population Inference}
\label{sec:sampling}

In this work, we sample from the posteriors of interest using a Hamiltonian Monte Carlo (HMC) algorithm implemented with \textsc{NumPyro} \citep{Bingham_2019, Phan_2019}. In particular, we use the No U-turn Sampler \citep[NUTS;][]{Hoffman_2011} variant of HMC from \textsc{NumPyro}. We use the Gelman-Rubin statistic, $\hat{R}$, to assess the convergence of the Markov Chains \citep{Gelman_Rubin_1992}. The sampling algorithms used in this work have been made available on GitHub at \url{https://github.com/erinhay/snia-zband}.

We retrieve the cosmological luminosity distance, $\mu_{\Lambda\text{CDM}}(z)$, from the redshift using functionality from \texttt{astropy} \citep{astropy:2013, astropy:2018, astropy:2022}. The redshift of each SN~Ia is corrected for peculiar velocities using the Cosmic Flows tools from \citet{Carrick_2015}. Throughout this work, we assume a peculiar velocity uncertainty of $\sigma_{\text{pec}} = 150 \, \text{km} \, \text{s}^{-1}$ \citep{Carrick_2015}. We additionally run the analysis with $\sigma_{\text{pec}} = 250 \, \text{km} \, \text{s}^{-1}$ and find that the the conclusions of the work are unchanged.

\subsubsection{Modelling the Absolute Magnitude Distribution}
\label{sec:sampling-absmag}
To understand how standard SNe~Ia are in the $z$ band relative to the $g$, $r$, and $i$ bands, we are interested in modelling the distribution of absolute magnitudes in each of these filters. Below, we describe our hierarchical Bayesian model for the population mean magnitude at peak in some generic band, as well as the population residual scatter around this mean value, for a sample of SNe~Ia.

We will first be interested in modelling the population mean ``dust-extinguished'' or ``extinguished,'' peak absolute magnitude, $M^{0}_{\text{ext}}$. Here, we use ``extinguished'' to indicate that the magnitudes have not been corrected for the effects of dust in the host galaxy. Second, we model the population mean peak absolute magnitude \textit{after} correction for host galaxy dust, which we refer to as the peak ``intrinsic'' absolute magnitude, $M^{0}_{\text{int}}$. We use the phrase ``at peak'' to denote ``at $T_{0}$.''

In our hierarchical Bayesian model, we assume that the latent peak extinguished/intrinsic absolute magnitude of SN~Ia $s$ is independently sampled from the population distribution:
\begin{equation}
    M^{s}_{\text{ext/int}} \sim \mathcal{N}(M^{0}_{\text{ext/int}}, \sigma^{2}_{\text{res, ext/int}}) \hspace{0.3cm}
\end{equation}
where $M^{0}_{\text{ext/int}}$ is the population mean peak extinguished/intrinsic absolute magnitude in some band, and $\sigma^{2}_{\text{res, ext/int}}$ is the variance around this population mean peak extinguished/intrinsic absolute magnitude. We use $\mathcal{N}(\mu, \sigma^{2})$ to denote a normal distribution with mean, $\mu$, and variance, $\sigma^{2}$.

We estimate the extinguished/intrinsic magnitude at peak for SN $s$ according to:
\begin{equation}
    \hat{M}^{s}_{\text{ext/int}} = \hat{m}^{s}_{\text{ext/int}} - \mu_{\Lambda\text{CDM}}(\hat{z}^{s})
\end{equation}
where $\hat{m}^{s}_{\text{ext/int}}$ is the extinguished/intrinsic apparent magnitude at peak inferred from \textsc{BayeSN} and $\mu_{\Lambda\text{CDM}}(\hat{z}^{s})$ is the cosmological luminosity distance evaluated at the CMB-frame redshift of the SN that has been corrected for bulk flows, $\hat{z}^{s}$.

The inferred extinguished/intrinsic magnitude at peak, $\hat{M}^{s}_{\text{ext/int}}$, is an uncertain estimate of the latent peak absolute magnitude, $M^{s}_{\text{ext/int}}$, for SN $s$. Therefore, the estimator of the peak magnitude can be related to the latent parameter according to:
\begin{equation}
    \hat{M}^{s}_{\text{ext/int}} \sim \mathcal{N}(M^{s}_{\text{ext/int}}, \sigma^{2}_{s})
    \label{eq:obs_M}
\end{equation}
where the uncertainty on the inferred peak magnitude is given by:
\begin{equation}
    \sigma_{s} = \sqrt{\sigma_{\text{fit}, s}^{2} + \sigma_{\text{cosmo}, s}^{2}}
    \label{eq:sigma_s}
\end{equation}
with $\sigma_{\text{fit}, s}$ denoting the uncertainty from the \textsc{BayeSN} fit for the apparent magnitude at peak of SN~Ia $s$, and $\sigma_{\text{cosmo}, s}$ denoting the uncertainty due to redshift-based cosmological distances.

The uncertainty due to redshift-based cosmological distances arises from uncertainty in observed redshift, $\hat{z}^{s}$, according to:
\begin{equation}
    \sigma_{\text{cosmo}, s}^{2} \approx \Big(\frac{5}{\hat{z}^{s} \text{ln}10}\Big)^{2} \; \Big[\sigma_{\text{pec}}^{2} / c^{2} + \sigma^{2}_{z,s}\Big]
\end{equation}
where $\sigma_{z,s}$ is the uncertainty on the measured redshift and $\sigma_{\text{pec}}$ is the uncertainty due to peculiar velocities. We assume the errors in fitting peak magnitudes, measuring redshifts, and the peculiar velocity corrections are all independent.

For the observed peak extinguished/intrinsic absolute magnitude of a single SN~Ia, $\hat{M}^{s}_{\text{ext/int}}$, the likelihood is given by,
\begin{equation}
    P(\hat{M}^{s}_{\text{ext/int}} | \, \bm{\Theta}, \sigma^{2}_{s}) = \mathcal{N}(\hat{M}^{s}_{\text{ext/int}}| \; M^{0}_{\text{ext/int}}, \sigma^{2}_{s} + \sigma_{\text{res, ext/int}}^{2})
\end{equation}
where $\bm{\Theta} = (M^{0}_{\text{ext/int}}, \sigma_{\text{res, ext/int}})$. The joint likelihood is thus given by:
\begin{equation}
    P(\hat{\bm{M}}_{\text{ext/int}} | \, \bm{\Theta}, \bm{\sigma}^{2}) = \prod_{s=1}^{N} P(\hat{M}^{s}_{\text{ext/int}} | \, \bm{\Theta}, \sigma^{2}_{s})
\end{equation}
for a sample of $N$ SNe~Ia.

The joint posterior is then given by:
\begin{equation}
    P(\bm{\Theta} | \, \bm{\hat{M}}_{\text{ext/int}}, \bm{\sigma}^{2}) \propto \Bigg[ \prod_{s=1}^{N} P(\hat{M}^{s}_{\text{ext/int}} | \, \bm{\Theta}, \sigma^{2}_{s}) \Bigg] P(\bm{\Theta})
\end{equation}
where we use the following hyperpriors on these hyperparameters:
\begin{gather}
    M^{0}_{\text{ext/int}} \sim \mathcal{U}(-\infty, \infty), \\
    \sigma_{\text{res, ext/int}} \sim \mathcal{U}(0, \infty)
\end{gather}
in the analysis presented in \S\ref{sec:analysis-stretchlum}. We use $\mathcal{U}(a, b)$ to denote a uniform distribution with lower bound $a$ and upper bound $b$.

\input{table2}
\subsubsection{Modelling the Shape-Luminosity Relation}
\label{sec:sampling-stretchlum}
Next, we explore the shape-luminosity relation -- the dependence of the peak intrinsic absolute magnitudes on the shape parameter $\theta$. We model the shape-luminosity relation with a simple linear relation, including a residual scatter term around this relation. We perform this analysis for the peak intrinsic absolute magnitudes in each band, which we compare to the shape parameter inferred from the $gri$ only \textsc{BayeSN} fit.

For a SN $s$ with a shape parameter, $\theta^{s}$, we model the dependence of the latent peak intrinsic absolute $z$-band magnitude on the shape parameter as:
\begin{equation}
    M^{s}_{z, \text{int}} | \, \hat{\theta}^{s} \sim \mathcal{N}( a \hat{\theta}^{s} + b, \sigma^{2}_{\text{res, S-L}})
\end{equation}
where $a$ and $b$ are the slope and intercept of the linear model and $\sigma^{2}_{\text{res, S-L}}$ is the residual scatter around the linear shape-luminosity model. We assume the latent peak intrinsic absolute magnitudes for a sample of SNe~Ia are conditionally independent given the shape parameters.

The observed peak magnitude is an uncertain estimate of the true peak magnitude:
\begin{equation}
    \hat{M}^{s}_{z, \text{int}} \sim \mathcal{N}(M^{s}_{z, \text{int}}, \sigma_{s}^{2})
\end{equation}
where the measurement uncertainty is given by Equation \ref{eq:sigma_s}. We neglect the measurement uncertainties on the shape parameter for each SN~Ia, as justified below.

The likelihood given the data for a single SN~Ia $s$ is thus:
\begin{equation}
    P(\hat{M}^{s}_{z, \text{int}} | \, \bm{\Theta}, \hat{\theta}^{s}, \sigma_{s}^{2}) = \mathcal{N}(\hat{M}^{s}_{z, \text{int}} | \, a\hat{\theta}^{s} + b, \sigma_{s}^{2} + \sigma_{\text{res, S-L}}^{2})
\end{equation}
where $\bm{\Theta} = (a, b, \sigma_{\text{res, S-L}})$ here. The joint likelihood for a sample of $N$ SNe~Ia is then:
\begin{equation}
    P(\hat{\bm{M}}_{z, \text{int}} | \, \bm{\Theta}, \hat{\bm{\theta}}, \bm{\sigma}^{2}) = \prod_{s=1}^{N} P(\hat{M}^{s}_{z, \text{int}} | \, \bm{\Theta}, \hat{\theta}^{s}, \sigma^{2}_{s})
\end{equation}
again assuming conditional independence given the hyperparameters of the model.

We can now sample the joint posterior according to:
\begin{equation}
    P(\bm{\Theta} | \, \bm{\hat{M}}_{z, \text{int}}, \hat{\bm{\theta}}, \bm{\sigma}^{2}) \propto \Bigg[ \prod_{s=1}^{N} P(\hat{M}^{s}_{z, \text{int}} | \, \bm{\Theta}, \hat{\theta}^{s}, \sigma^{2}_{s}) \Bigg] P(\bm{\Theta})
\end{equation}
where we will use the following hyperpriors on the shape-luminosity relation hyperparameters:
\begin{gather}
    a \sim \mathcal{U}(-\infty, \infty), \\
    b \sim \mathcal{U}(-\infty, \infty), \\
    \sigma_{\text{res, S-L}} \sim \mathcal{U}(0, \infty)
\end{gather}
when presenting the results of this analysis in \S\ref{sec:analysis-stretchlum}.

Note that we are are regressing against the measured shape parameters, rather than the latent shape parameters. Neglecting measurement uncertainties on shape, the independent variable, can lead to regression dilution \citep{Kelly_2007, Kelly_2012}, whereby the slope in a linear model becomes biased towards zero due to mistreatment of uncertainties on the independent variable. We use forward simulations to test that the bias on the estimated slope of the shape-luminosity relation, $a$, is small in this case. For this test, we regress the peak $z$-band magnitudes against $\theta_{gri}$. In our forward simulations, we assume that the shape parameter is drawn from a population unit normal distribution. Each latent shape parameter is assumed to be measured with a Gaussian error, with measurement uncertainty based on the distribution of $\theta$ uncertainties from \textsc{BayeSN} for the real data.

To determine realistic values for the shape-luminosity hyperparameters, we first fit the real data with a model that neglects measurement uncertainties on the shape parameter, as shown in Figure \ref{fig:stretch-lum}. Informed by this result, we assume $a=0.15$, $b=-18.3$ mag, and $\sigma_{\text{res, S-L}}=0.15$. We then simulate 100 realisations of the data and fit each realisation with our model, which neglects measurement error on $\theta$. For each realisation of the data, we retrieve the fitted shape-luminosity slope by taking the posterior mean of $a$. We then examine the distribution of fitted slopes for the 100 realisations of simulated data. The sample mean of the distribution of fitted slopes is only slightly offset by $-0.006$ relative to the true value of $a=0.15$. The conclusions of this work are insensitive to a $\sim4\%$ level of bias. 

\subsubsection{Modelling the Mass Step}
\label{sec:sampling-massstep}
We are also interested in estimating the size of the mass step in the sample. The mass step is a difference in the population mean Hubble residual of SNe which explode in high mass (HM) host galaxies and those which explode in low mass (LM) host galaxies.

When estimating the mass step, we use the fitted distance modulus from \textsc{BayeSN}, rather than the peak magnitudes. The reason for this choice is: 1) the distance modulus from \textsc{BayeSN} is corrected for the shape-luminosity relation comprehensively learned from $griz$ training data, rather than the individual-band linear model adopted in \S\ref{sec:sampling-stretchlum}, and 2) it allows for a like-for-like comparison between a fit to data in multiple bands (e.g., a fit to the optical $gri$ data) and a fit to data in an individual band.

Recall that the Hubble residual of SN~Ia $s$ is given by $\hat{\delta}^{s} = \hat{\mu}^{s} - \mu_{\Lambda\text{CDM}}(\hat{z}^{s})$, where $\hat{\mu}^{s}$ is the photometric distance modulus inferred from the \textsc{BayeSN} fit and $\mu_{\Lambda\text{CDM}}(\hat{z}^{s})$ is the cosmological distance modulus from the measured redshift of the SN, $\hat{z}^{s}$. Assuming that the mass step is a step function which transitions at $\mathcal{M}_{\text{step}}$, the Hubble residual of SN~Ia $s$ will depend on the mass of the SN's host galaxy, $\mathcal{M}^{s}_{*}$, according to:
\begin{gather}
    \delta^{s} | \, \mathcal{M}^{s}_{*} \sim
    \begin{cases}
        \mathcal{N}(\delta_{\text{LM}}, \sigma_{\text{res, LM}}^{2}) & \text{if} \; \mathcal{M}^{s}_{*} < \mathcal{M}_{\text{step}} \\
        \mathcal{N}(\delta_{\text{HM}}, \sigma_{\text{res, HM}}^{2}) & \text{if} \; \mathcal{M}^{s}_{*} \geq \mathcal{M}_{\text{step}}
    \end{cases}
    \label{eq:mass-step}
\end{gather}
where $\delta_{\text{LM/HM}}$ is the population mean Hubble residual in the LM/HM host galaxy bin and $\sigma_{\text{res, LM/HM}}^{2}$ is the residual population scatter of SNe absolute magnitudes in the LM/HM host galaxy bins. We assume that in each mass bin, the latent Hubble residual for each SN is independently and identically drawn according to Equation \ref{eq:mass-step}.

The inferred Hubble residual, which is an uncertain estimate of the true Hubble residual, is given by:
\begin{equation}
    \hat{\delta}^{s} \sim \mathcal{N}(\delta^{s}, \sigma_{s}^{2})
\end{equation}
with some measurement uncertainty on the Hubble residual, given by Equation \ref{eq:sigma_s} where $\sigma_{\text{fit}, s}^{2}$ now comes from the uncertainty on the inferred distance modulus, $\hat{\mu}^{s}$, from \textsc{BayeSN}. Again we assume independence of all measurement errors.

For each mass bin, there are two population-level parameters to infer: $\bm{\Theta} = (\delta_{\text{LM/HM}}, \sigma^{2}_{\text{res, LM/HM}})$. The likelihood in each mass bin is therefore given by:
\begin{equation}
    P(\hat{\delta}^{s} | \, \bm{\Theta}, \mathcal{M}^{s}_{*}, \sigma_{s}^{2}) = \mathcal{N}(\hat{\delta}^{s} | \, \delta_{\text{LM/HM}}, \sigma_{s}^{2} + \sigma_{\text{res, LM/HM}}^{2})
\end{equation}
for some SN~Ia $s$.

We define the joint likelihood for a sample of $N$ SNe~Ia to be:
\begin{equation}
    P(\bm{\hat{\delta}} | \, \bm{\Theta}, \bm{\mathcal{M}_{*}}, \bm{\sigma}^{2}) = \prod_{s=1}^{N} P(\hat{\delta}^{s} | \, \bm{\Theta}, \mathcal{M}^{s}_{*}, \sigma_{s}^{2})
\end{equation}
which we use to define the joint posterior:
\begin{equation}
    P(\bm{\Theta} | \, \bm{\hat{\delta}}, \bm{\mathcal{M}_{*}}, \bm{\sigma}^{2}) \propto \Bigg[ \prod_{s=1}^{N} P(\hat{\delta}^{s} | \, \bm{\Theta}, \mathcal{M}^{s}_{*}, \sigma_{s}^{2}) \Bigg] P(\bm{\Theta})
\end{equation}
where we adopt the following uninformative hyperpriors on the hyperparameters:
\begin{gather}
    \delta_{\text{LM}} \sim \mathcal{U}(-\infty, \infty), \\
    \delta_{\text{HM}} \sim \mathcal{U}(-\infty, \infty), \\
    \sigma_{\text{res, LM}} \sim \mathcal{U}(0, \infty), \\
    \sigma_{\text{res, HM}} \sim \mathcal{U}(0, \infty)
\end{gather}
for the analysis in \S\ref{sec:analysis-massstep}. To compute a posterior on the mass step, $\gamma = \delta_{\text{HM}} - \delta_{\text{LM}}$, we take the difference of the population mean Hubble residual in each mass bin for each step along the chain.

To account for measurement uncertainties on the host galaxy mass estimates, we take 100 samples from the host mass posteriors from \texttt{Blast} for each object and re-run the inference for each of these realisations of the data. We take the variance of the 100 resulting mass steps that we infer as an additional uncertainty that we add in quadrature with the uncertainty on the mass step inferred when using the posterior median host mass for each object. The contribution to the uncertainty is $0.008$ mag on average.

%% file: table2.tex
\begin{table*}
    \centering
    \caption{Properties of the peak magnitudes of SNe~Ia in each of the $griz$ filters. We report the residual scatter in the extinguished peak magnitudes, $\sigma_{\text{res, ext}}$, the population mean peak intrinsic absolute magnitude, $M^{0}_{\text{int}}$, and the residual scatter around the mean peak intrinsic absolute magnitude, $\sigma_{\text{res, int}}$. For the $z$-band, we also report the slope of the shape-luminosity relation, $a$, and the residual scatter around the shape-luminosity relation, $\sigma_{\text{res, S-L}}$.}
    \begin{tabular}{c|c|c|c|c|c}
        Filter & $\sigma_{\text{res, ext}}$ [mag] & $M^{0}_{\text{int}}$ [mag] & $\sigma_{\text{res, int}}$ [mag] & $a$ & $\sigma_{\text{res, S-L}}$ \\
        \hline
        \hline
        $g$ & $\gExtResScat \pm \gExtResScatErr$ & $\gPeakIntAbsMag \pm \gPeakIntAbsMagErr$ & $\gIntResScat \pm \gIntResScatErr$ & $\gStretchLumSlope \pm \gStretchLumSlopeErr$ & $\gStretchLumScat \pm \gStretchLumScatErr$ \\
        $r$ & $\rExtResScat \pm \rExtResScatErr$ & $\rPeakIntAbsMag \pm \rPeakIntAbsMagErr$ & $\rIntResScat \pm \rIntResScatErr$ & $\rStretchLumSlope \pm \rStretchLumSlopeErr$ & $\rStretchLumScat \pm \rStretchLumScatErr$ \\
        $i$ & $\iExtResScat \pm \iExtResScatErr$ & $\iPeakIntAbsMag \pm \iPeakIntAbsMagErr$ & $\iIntResScat \pm \iIntResScatErr$ & $\iStretchLumSlope \pm \iStretchLumSlopeErr$ & $\iStretchLumScat \pm \iStretchLumScatErr$ \\
        $z$ & $\zExtResScat \pm \zExtResScatErr$ & $\zPeakIntAbsMag \pm \zPeakIntAbsMagErr$ & $\zIntResScat \pm \zIntResScatErr$ & $\zStretchLumSlope \pm \zStretchLumSlopeErr$ & $\zStretchLumScat \pm \zStretchLumScatErr$ \\
        \hline
    \end{tabular}
    \label{tab:peak-mags-table}
\end{table*}

%% file: 04_analysis.tex
\subsection{Standardisation of SNe~Ia Peak z-band Magnitudes}
\label{sec:analysis-stretchlum}
To assess the utility of SNe~Ia before applying empirical light curve corrections, we can estimate the population mean peak extinguished absolute magnitude and residual population scatter using the apparent magnitudes as extracted from \textsc{BayeSN}. We will carry out this analysis for each of the individual bands according to the methodology presented in \S\ref{sec:sampling-absmag}.

In Table \ref{tab:peak-mags-table}, we report the residual scatter around the peak extinguished absolute magnitude by band found using \S\ref{sec:sampling-stretchlum}. For the $g$, $r$, and $i$ bands, we find $\sigma_{\text{res, ext}}(g) = \gExtResScat \pm \gExtResScatErr$ mag, $\sigma_{\text{res, ext}}(r) = \rExtResScat \pm \rExtResScatErr$ mag, and $\sigma_{\text{res, ext}}(i) = \iExtResScat \pm \iExtResScatErr$ mag, respectively. In the $z$ band, we find $\sigma_{\text{res, ext}}(z) = \zExtResScat \pm \zExtResScatErr$ mag. These results demonstrate that the redder $i$ and $z$ bands benefit from being relatively better standard candles than the optical $g$ and $r$ bands. This finding is likely contributed to by the greater transparency of the redder wavelength regime to dust, in addition to a decreased intrinsic scatter in the $i$ and $z$ bands.

\begin{figure}
    \centering
    \includegraphics[width=1\linewidth]{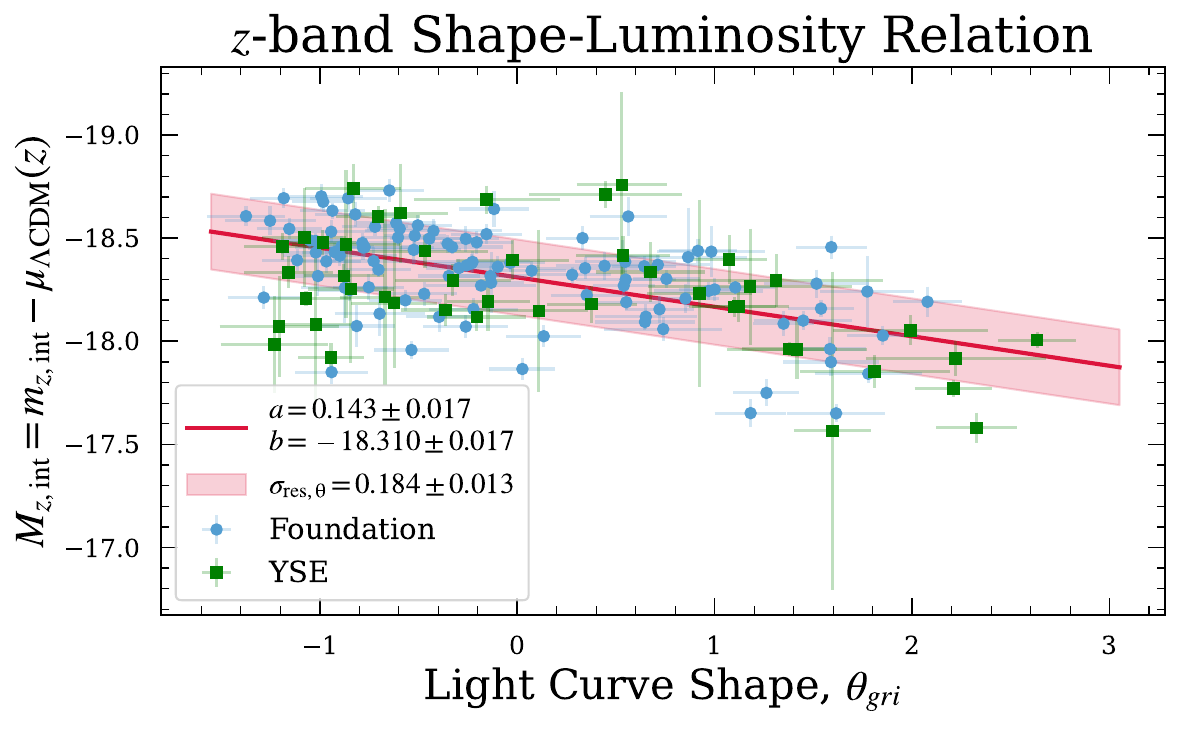}
    \caption{The $z$-band peak intrinsic absolute magnitude, $M_{z, \text{int}} = m_{z, \text{int}} - \mu_{\Lambda \text{CDM}}(z)$, as a function of the inferred shape parameter from a \textsc{BayeSN} fit to optical $gri$ data only, $\theta_{gri}$. The fitted shape-luminosity relation is shown as a solid line. The shaded region shows the fitted residual scatter around this relation.}
    \label{fig:stretch-lum}
\end{figure}

\begin{figure*}
    \centering
    \includegraphics[width=\linewidth]{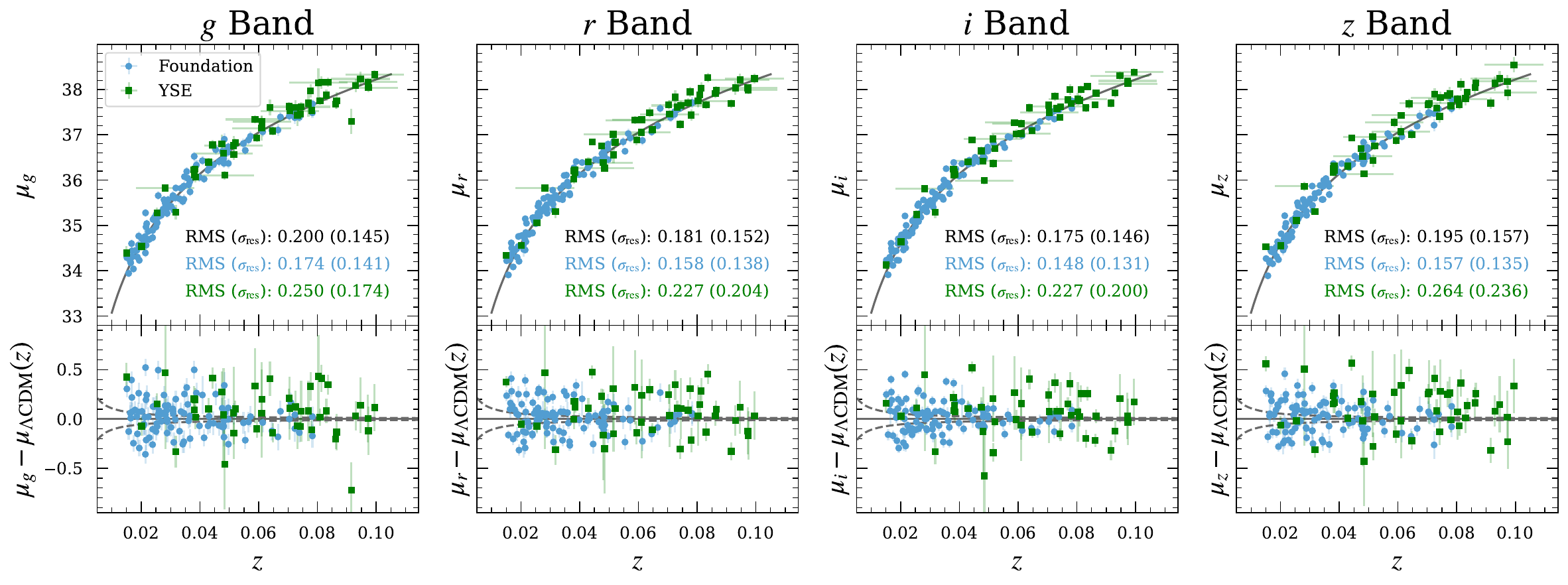}
    \caption{The Hubble diagrams from the corrected distance modulus from the individual-band \textsc{BayeSN} fits. From left to right, the Hubble diagrams are constructed from the $g$-band, $r$-band, $i$-band, and $z$-band distances, respectively. All individual-band Hubble diagrams are corrected for dust assuming $A_{V}$ from the fit to the $griz$ data. In the Hubble diagrams (top row), the x-axis error bars represent the redshift measurement uncertainties and the y-axis error bars represent \textsc{BayeSN} distance modulus fitting uncertainties (not including residual scatter, $\sigma_{\text{res}}$). In the Hubble residual plots (bottom row), the redshift measurement uncertainties have been propagated to an error on the residual. The dashed lines in the Hubble residual plots show the level of distance modulus uncertainty that can be expected from peculiar velocities assuming $\sigma_{\text{pec}} = 150 \, \text{km} \, \text{s}^{-1}$. The y-axis of all subplots have units of magnitude.}
    \label{fig:distance-modulus-hubble-diagram-single-band}
\end{figure*}

\begin{figure}
    \centering
    \includegraphics[width=\linewidth]{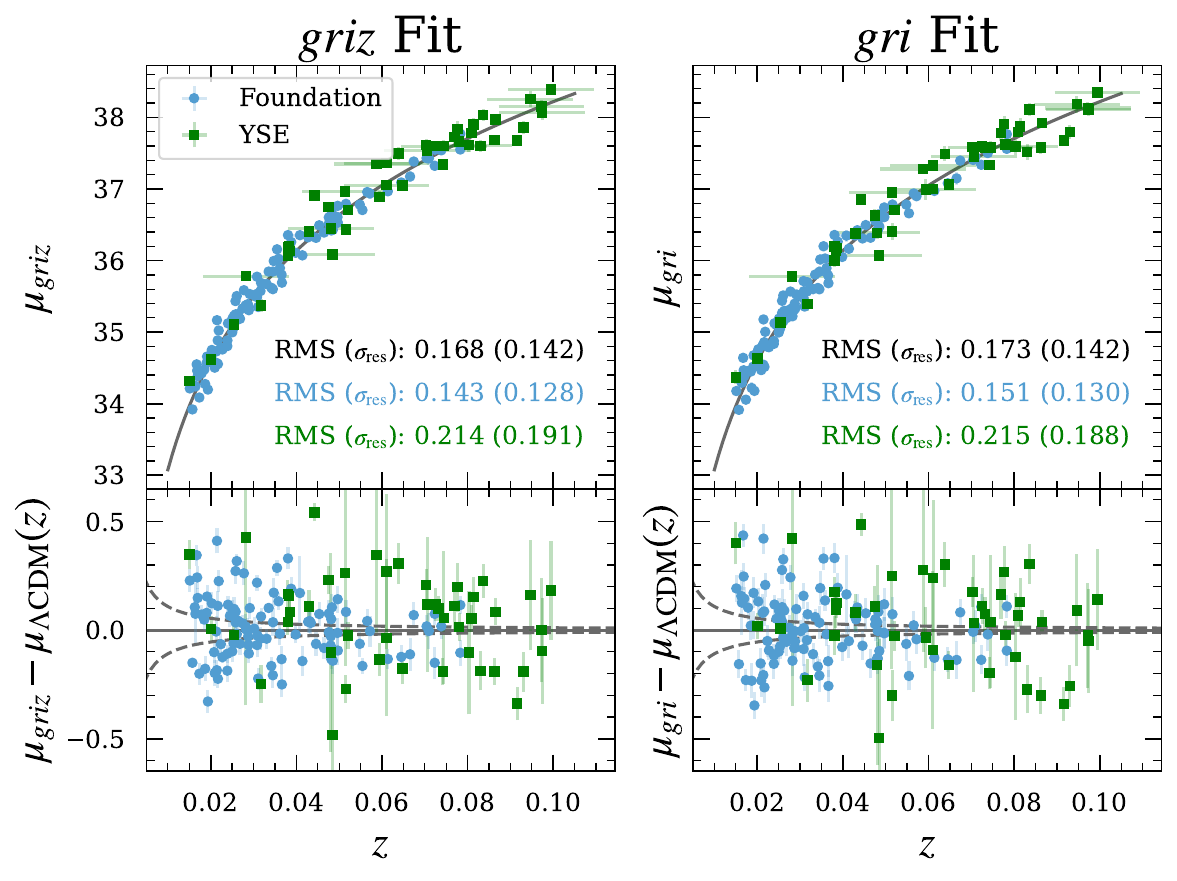}
    \caption{The Hubble diagrams from the corrected distance modulus from the multi-band \textsc{BayeSN} fits. (Left) The Hubble diagram using the distance moduli from the fit to the full $griz$ light curves. The Hubble residuals have RMS = $\grizRMS$ mag. (Right) The Hubble diagram and residuals from the fit to the optical only $gri$ light curves. The residuals from this fit have RMS = $\griRMS$ mag. The uncertainties along each axis are as described in Figure \ref{fig:distance-modulus-hubble-diagram-single-band}.}
    \label{fig:distance-modulus-hubble-diagram-multi-band}
\end{figure}

Next, we correct the peak extinguished absolute magnitudes for dust extinction to yield the peak intrinsic absolute magnitudes. We report the population mean intrinsic absolute magnitude and residual scatter around the mean for each band in Table \ref{tab:peak-mags-table}. In the $g$-band, we find $\sigma_{\text{res, int}}(g) = \gIntResScat \pm \gIntResScatErr$ mag, in the $r$-band, we find $\sigma_{\text{res, int}}(r) = \rIntResScat \pm \rIntResScatErr$ mag, and in the $i$-band, we find $\sigma_{\text{res, int}}(i) = \iIntResScat \pm \iIntResScatErr$ mag. We find the residual scatter in the $z$ band to be $\sigma_{\text{res, int}}(z) = \zIntResScat \pm \zIntResScatErr$ mag. The residual scatter around the peak intrinsic absolute magnitude remains higher for the $g$ and $r$ bands than for the $i$ and $z$ bands. Therefore, SNe~Ia appear to be relatively better standard candles the $i$ and $z$ band than in the $g$- and $r$-band. Because of the $z$ band's greater transparency to dust, it shows only a small reduction in the residual scatter from peak magnitudes pre-dust correction to post-dust correction, as anticipated.

We next investigate the dependence of the $z$-band peak intrinsic magnitudes to the inferred shape parameter, $\theta$ from the $gri$ fit to the data, according to the methodology in \S\ref{sec:sampling-stretchlum}. We find the slope of the shape-luminosity relation to be $a = \zStretchLumSlope \pm \zStretchLumSlopeErr$ with a residual scatter around the fitted linear relation of $\sigma_{\text{res, S-L}} = \zStretchLumScat \pm \zStretchLumScatErr$ mag. Figure \ref{fig:stretch-lum} shows $M_{z, \text{int}}$ versus $\theta_{gri}$ with the fitted linear shape-luminosity relation. There is a statistically significant correlation between the peak intrinsic $z$-band magnitude and the shape of the light curve, as the estimated shape-luminosity slope is inconsistent with zero at the $8.4\sigma$ level. 

\subsection{The Hubble Diagram}
\label{sec:analysis-hubblediagram}
We infer the scatter in the Hubble residuals from the corrected distance moduli to the SNe~Ia. The Hubble diagrams and residuals from the individual-band \textsc{BayeSN} fits are shown in Figure \ref{fig:distance-modulus-hubble-diagram-single-band}. In Figure \ref{fig:distance-modulus-hubble-diagram-multi-band}, we additionally show the $griz$ and $gri$ Hubble diagrams. In all Hubble diagrams, the mean Hubble residuals are consistent with zero, indicating good calibration of the model. We use the root mean square (RMS) of the Hubble residuals to estimate the total scatter in the Hubble residuals, as this metric is insensitive to modelling assumptions (e.g., assumed peculiar velocity uncertainty). To give a measure of the residual scatter that is unexplained by measurement uncertainties and peculiar velocities, we estimate the residual scatter using the methodology of \S\ref{sec:sampling-absmag} applied to Hubble residuals. 

The individual-band Hubble diagrams exhibit a total scatter of RMS $= \gRMS$ mag in the $g$ band, RMS $= \rRMS$ mag in the $r$ band, RMS $= \iRMS$ mag in the $i$ band, and RMS $= \zRMS$ mag in the $z$ band. Interestingly, the RMS decreases with increasing wavelength from the $g$ to $i$ bands, and then increases in the $z$ band. The increase in scatter in the $z$-band Hubble diagram appears to be driven by the YSE subset of the sample. Therefore, we attribute some of this increase to the lower median number of $z$-band observations (4 observations per object) compared to in the other filters (between 6-7 observations per object) in the sample. The Foundation subset of the sample, which has a more balanced number of observations in each bands, shows trends similar to those seen in the peak magnitude scatter in \S\ref{sec:analysis-stretchlum} where the $z$-band peak magnitudes show a level of scatter that falls between the $r$- and $i$-band peak magnitudes. Based on the Hubble diagrams in this work, we can conclude that SNe~Ia in the $z$ band are relatively worse standard candles than in the $i$ band, and their relative standardisation compared to the $g$ and $r$ bands cannot be definitely determined from this sample of SNe~Ia.

Analysing the residual scatter, $\sigma_{\text{res}}$, of the Hubble residuals from the combined sample and individual surveys leads to similar conclusions. The residual scatter estimates for the combined sample are $\sigma_{\text{res}}(g) = \gsigmares$ mag, $\sigma_{\text{res}}(r) = \rsigmares$ mag, $\sigma_{\text{res}}(i) = \isigmares$ mag, and $\sigma_{\text{res}}(z) = \zsigmares$ mag. When considering the Foundation sample on its own, the $z$ band again exhibits a level of scatter that is slightly reduced compared to the $r$ band, but slightly increased relative to the $i$ band. The YSE sample does not show as clear a trend, with the $g$-band Hubble diagram having the lowest residual scatter and the $z$-band Hubble diagram having the highest residual scatter.

We also compute an inverse variance weighted RMS (wRMS), where the variance includes uncertainties from cosmological based distances, redshift measurement, and the \textsc{BayeSN} fit, as well as residual scatter. The residual scatter for each Hubble diagram is taken from the $\sigma_{\text{res}}$ values reported above. We find wRMS $=\gwRMS$ mag in the $g$ band, wRMS $=\rwRMS$ mag in the $r$ band, wRMS $=\iwRMS$ mag in the $i$ band, and wRMS $=\zwRMS$ mag in the $z$ band. These results lead to consistent conclusions with those drawn from the unweighted RMS.

When comparing the $griz$ and $gri$ Hubble diagrams, we find that the RMS decreases from RMS $= \griRMS$ mag in the $gri$ Hubble diagram to RMS $= \grizRMS$ mag in the $griz$ Hubble diagram. The scatter in the optical $gri$ Hubble diagram is lower than the $z$-band Hubble diagram, though this result is not unexpected given differences in the amount of data going into each fit. Importantly, we find that combining the $z$-band data with the optical does improve distances estimates.

Again, the conclusions from the analysis of the residual scatter are not as clear cut, with both the $griz$ and $gri$ Hubble diagrams having $\sigma_{\text{res}} = 0.142$ mag. This finding is also driven by the YSE subset of the sample, as Foundation does show a decrease in residual scatter when including the $z$-band data; we comment on the differences between YSE and Foundation which may lead to this result in the following paragraph. Finally, the weighted RMS of the $griz$ Hubble diagram is wRMS $= \grizwRMS$ mag compared to the relatively higher weighted RMS from the $gri$ Hubble diagram of wRMS $= \griwRMS$ mag, which is a consistent result as from the unweighted RMS.

Further to the differences between the two surveys, the total and residual scatter of the YSE subset is consistently higher than that of the Foundation subset. We attribute this discrepancy to the small number of YSE objects which pass the requisite data cuts, which makes the total scatter more sensitive to outliers, the larger number of objects with redshifts derived from the SN~Ia spectra, for which the assumed measurement uncertainty of $\sigma_{z} = 0.01$ dominates the error budget, and the lower cadence of the survey relative to Foundation. Indeed, the median number of $z$-band observations is five for objects from Foundation and two for objects from YSE, before data cuts. The reason for this difference can largely be attributed to YSE being an untargeted survey, unlike Foundation, so the survey scans the sky according to a predetermined schedule. The selection effects are better defined in untargeted surveys, which is beneficial in cosmological analyses, but there are trade-offs in the cadence of the light curves from YSE DR1. While we mitigate some of these differences with the requirement that there be four $z$-band observations, there are still differences in the resulting $z$-band light curve data of the two surveys.

\subsection{The Mass Step}
\label{sec:analysis-massstep}
Finally, we estimate the mass step according to the methodology presented in \S\ref{sec:sampling-massstep} using the Hubble residuals from each of the $griz$, $gri$, and individual-band fits.

\input{table3}

Notably, we include separate residual scatter terms for the Foundation and YSE subsets of the sample. When fitting the mass step separately for each survey, we find that the SNe~Ia from the two surveys exhibit significantly different residual scatter properties. This finding is consistent with the trends in Hubble residual scatter as presented in \S\ref{sec:analysis-hubblediagram}. We modify Equation \ref{eq:mass-step} so that $\sigma^{2}_{\text{res, LM/HM}}$ is dependent on whether SN $s$ is from Foundation or YSE. We still fit for one mass step for both samples, as we find that the mass step remains consistent when fit separately for each survey.

For our analysis, we compare two choices of the mass step location. Firstly, we use $\mathcal{M}_{\text{step}} = 10^{10} \mathcal{M}_{\odot}$, as this value has been used in numerous previous analyses of the mass step \citep{Sullivan_2010, Jones_2019, Johansson_2021, Thorp_2021, TM_2022, Brout_2022, Grayling_2024}. This choice places \NlowTen objects in the low mass bin and \NhighTen objects in the high mass bin. Secondly, we use $\mathcal{M}_{\text{step}} = 10^{\medianhostgalmass} \mathcal{M}_{\odot}$, the median host galaxy mass in the sample. With this choice of mass step location, we get a more balanced split of the sample with \NMedian objects in each of the low and high mass bins. 

The results of the analysis on the combined YSE and Foundation sample, and for each survey individually, are presented for both mass step locations under the fiducial assumption of a single $R_{V}$ for all SNe~Ia in Table \ref{tab:hubble-table-singlerv}. With $\mathcal{M}_{\text{step}} = 10^{10} \mathcal{M}_{\odot}$, we find $\gamma_{griz} = \grizMassStepTen \pm \grizMassStepErrTen$ mag for the Hubble residuals from the distance moduli from the $griz$ data. The reported uncertainty includes a sub-dominant contribution arising from uncertainty in the host galaxy mass estimates. The residual scatter in the LM bin, $\sigma_{\text{res, LM}} = \grizYSELMScatTen \pm \grizYSELMScatErrTen$ mag for YSE and $\sigma_{\text{res, LM}} = \grizFoundationLMScatTen \pm \grizFoundationLMScatErrTen$ mag for Foundation. In the HM bin, we find $\sigma_{\text{res, HM}} = \grizYSEHMScatTen \pm \grizYSEHMScatErrTen$ mag for YSE and $\sigma_{\text{res, HM}} = \grizFoundationHMScatTen \pm \grizFoundationHMScatErrTen$ mag for Foundation. As in \S\ref{sec:analysis-hubblediagram}, the residual scatter in the YSE subset of the sample is consistently larger than the residual scatter in the Foundation subset of the sample.

\begin{figure}
    \centering
    \includegraphics[width=0.95\linewidth]{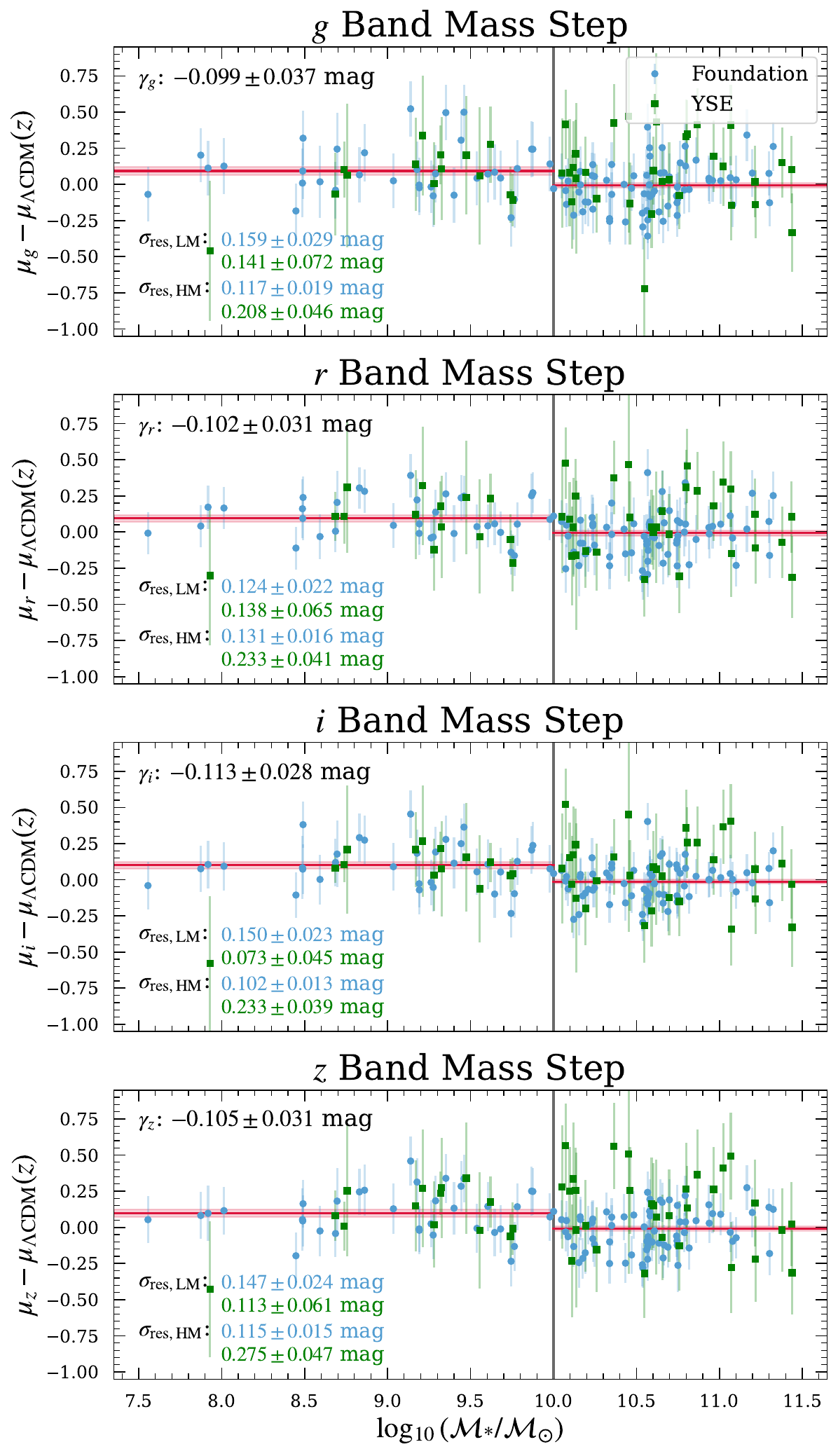}
    \caption{The Hubble residuals versus host galaxy mass for the four Hubble diagrams shown in Figure \ref{fig:distance-modulus-hubble-diagram-single-band} (From top to bottom: $g$-band, $r$-band, $i$-band, $z$-band). All distances have been corrected for host galaxy dust extinction assuming $R_{V} = 2.61$, assuming $A_{V}$ as fit from the $griz$ light curve. The y-axis of all subplots have units of magnitude. The horizontal red lines on either side of the mass step location ($\mathcal{M}_{\text{step}} = 10^{10} \mathcal{M}_{\odot}$, marked with a solid grey vertical line) show the population mean Hubble residual for the LM and HM bins, with the uncertainty on the population mean shown as the red shaded region. The inferred mass step is reported in the upper left hand corner of each subplot. The residual scatter in each mass bin for both surveys is shown in the lower left hand corner of each subplot. For visual clarity, we do not show the host galaxy mass uncertainties.}
    \label{fig:mass-step}
\end{figure}

\input{table4}

\begin{figure*}
    \centering
    \includegraphics[width=\linewidth]{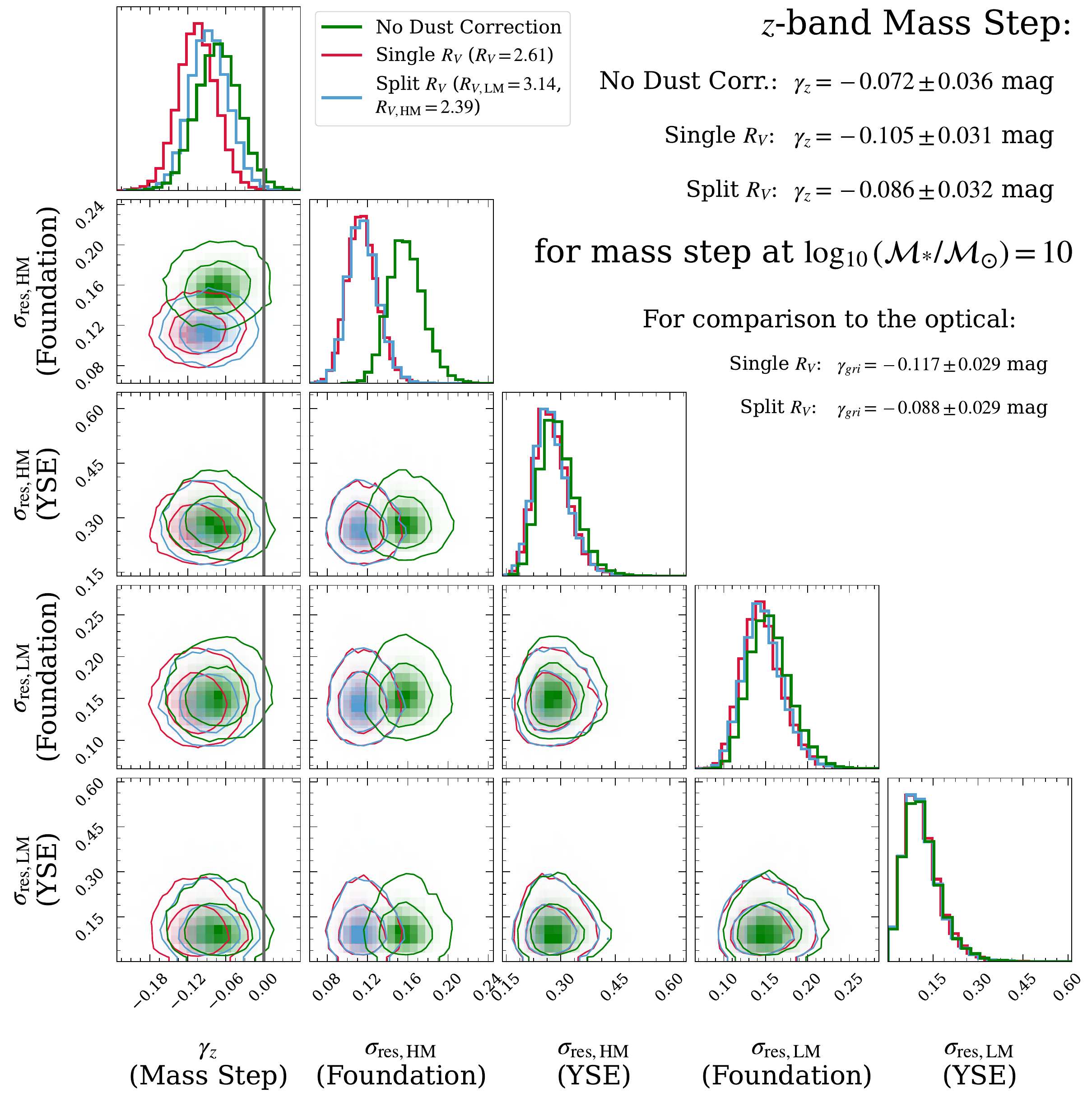}
    \caption{The posterior for the inferred mass step parameters from the fit to the $z$-band distance modulus that has not been corrected for host galaxy dust extinction (green), that has been corrected for host galaxy dust extinction assuming a single $R_{V}$ for all SNe~Ia (red), and that has been corrected for extinction assuming $R_{V}$ is different for SNe~Ia in high mass vs low mass host galaxies (blue). For the split $R_{V}$ case, we use $R_{V, \text{LM}} = 3.14$ for SNe~Ia in low mass host galaxies and $R_{V, \text{HM}} = 2.39$ for SNe~Ia in high mass host galaxies, according to the results in \citet{Grayling_2024}. These results assume a mass step location of $10^{10} \mathcal{M}_{\odot}$. For comparison, we also report the mean and standard deviation of the posterior samples of the mass step estimated from the optical $gri$ data under the assumption of a single vs split $R_{V}$.}
    \label{fig:mass-step-corner}
\end{figure*}

In addition, we estimate the mass step in individual filters and find $\gamma_{g} = \gMassStepTen \pm \gMassStepErrTen$ mag, $\gamma_{r} = \rMassStepTen \pm \rMassStepErrTen$ mag, $\gamma_{i} = \iMassStepTen \pm \iMassStepErrTen$ mag, and $\gamma_{z} = \zMassStepTen \pm \zMassStepErrTen$ mag. The mass steps estimated in each wavelength regime are all consistent within $1\sigma$ of each other, and of the $griz$ mass step. The residual scatter for each of these mass step estimates is also consistent within $1\sigma$ of that measured for the $griz$ mass step in each mass bin. Therefore, we estimate a mass step at $3$--$4.5\sigma$ significance that is consistent from the optical $gri$ bands to the $z$ band.

When considering a step at the median host galaxy mass of the sample, we find $\gamma_{griz} = \grizMassStepMedian \pm \grizMassStepErrMedian$ mag. Individually in each band we find $\gamma_{g} = \gMassStepMedian \pm \gMassStepErrMedian$ mag, $\gamma_{r} = \rMassStepMedian \pm \rMassStepErrMedian$ mag, $\gamma_{i} = \iMassStepMedian \pm \iMassStepErrMedian$ mag, and $\gamma_{z} = \zMassStepMedian \pm \zMassStepErrMedian$ mag. Again, the residual scatter terms are broadly consistent in trend with those estimated for $\mathcal{M}_{\text{step}} = 10^{10} \mathcal{M}_{\odot}$. Our results are slightly less significant for this choice of mass step location. Still, there is evidence for a non-zero mass step from the optical to the $z$ band that persists for both choices of mass step location. The results are also consistent when considering the Foundation and YSE subsets of the sample individually.

Figure \ref{fig:mass-step} visualises the results by individual filter assuming a mass step location of $10^{10} \mathcal{M}_{\odot}$. Shown is the host galaxy mass versus Hubble residual for the Hubble residuals from the four Hubble diagrams constructed from $\mu_{g}$, $\mu_{r}$, $\mu_{i}$, and $\mu_{z}$. Recall that these four sets of Hubble residuals have been corrected for dust in post-processing assuming $A_{V}$ from the $griz$ fit and $R_{V} = 2.61$. The fitted population mean Hubble residual in the low and high mass host bins, $\delta_{\text{LM/HM}}$ is also plotted with $1\sigma$ uncertainties. In addition, the residual scatter in each mass bin and for each survey are reported in the figure.

We additionally fit for a mass step assuming different values of $R_{V}$ for SNe~Ia in low mass host galaxies and for those in high mass host galaxies, rather than the fiducial assumption of a single $R_{V}$ for all SNe~Ia. For the split $R_{V}$ case, we assume $R_{V, \text{LM}} = 3.14$ for SNe~Ia in low mass host galaxies and $R_{V, \text{HM}} = 2.39$ for SNe~Ia in high mass host galaxies, according to the results from \citet{Grayling_2024}. We use the results from the analysis in \citet{Grayling_2024} that rely only on dust to explain differences in SNe~Ia in high mass and low mass host galaxies. The results of this analysis are reported in full in Table \ref{tab:hubble-table-splitrv}. Figure \ref{fig:mass-step-corner} compares the posterior on the mass step parameters assuming a single $R_{V}$ to that assuming a split on $R_{V}$ based on host galaxy mass for a step location of $\mathcal{M}_{\text{step}} = 10^{10} \mathcal{M}_{\odot}$. For this mass step location, we estimate $\gamma_{z} = \zSplitMassStepTen \pm \zSplitMassStepErrTen$ mag for the split $R_{V}$ case. For $\mathcal{M}_{\text{step}} = 10^{\medianhostgalmass} \mathcal{M}_{\odot}$, we estimate $\gamma_{z} = \zSplitMassStepMedian \pm \zSplitMassStepErrMedian$ mag for the split $R_{V}$ case.

Notably, the mass step in the $z$ band is reduced when assuming a split $R_{V}$ compared to the single $R_{V}$ case. The mass step in the optical is also reduced in this case compared to when assuming a single $R_{V}$. To contextualise this result, we highlight that the sample average extinction in the $z$ band when assuming $R_{V} = 2.61$, $\langle A_{z, R_{V} = 2.61} \rangle = 0.096$ mag. For this choice of $R_{V}$, the $z$ band extinction is $2.60\times$ less than the extinction in the $g$ band. When assuming $R_{V} = 3.14$ in the low mass bin, the average $z$-band extinction of the objects in the low mass bin is larger by $0.014$ mag than the average extinction in that bin when assuming $R_{V} = 2.61$. For the high mass bin, the average $z$-band extinction when assuming $R_{V}=2.39$ decreases by $0.009$ mag from the average in the high mass bin in the single $R_{V}$ case. These magnitude offsets explain the reduction in the size of the mass step when changing one's assumption from the single $R_{V}$ to the split $R_{V}$ case, and suggests dust plays some role in the mass step. That we still estimate a consistent, non-zero mass step across wavelength in the split $R_{V}$ case, though, suggests that dust alone cannot explain the mass step.

Finally, we test the effect of not applying a dust correction to the individual-band distance estimates on the mass step inference. In this analysis, we are agnostic to any assumptions about the dust law shape in different host galaxies. Because of the large scatter in the dust-extinguished distances, the optical mass steps are not well constrained in this analysis variant. For the same reason, the $z$-band mass step is found to be of lower significance than under the single $R_{V}$ and split $R_{V}$ assumptions about dust. We compare the ``No Dust Correction'' posterior on the $z$-band mass step parameters to that of the single $R_{V}$ and split $R_{V}$ cases in Figure \ref{fig:mass-step-corner}.

\begin{figure}
    \centering
    \includegraphics[width=\linewidth]{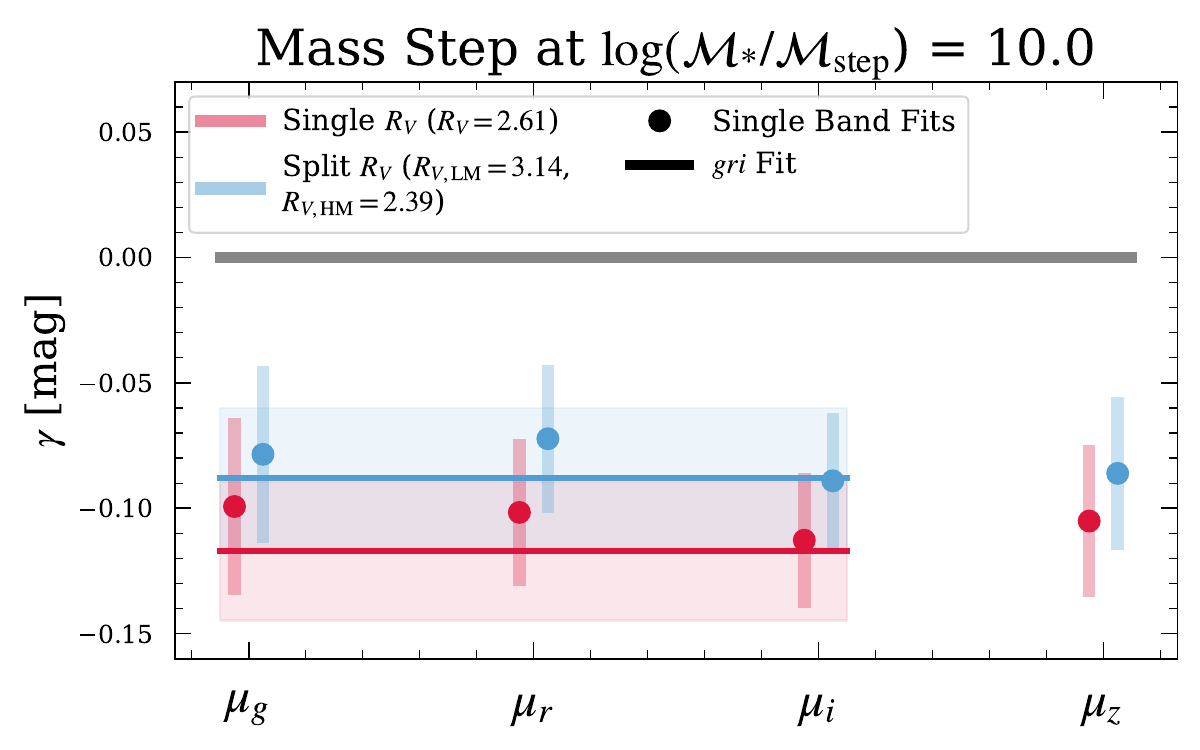}
    \caption{The mass step as a function of filter. The points represent individual-band mass steps, from the \textsc{BayeSN} fits to the individual-band light curves. The solid lines represent the $gri$ mass step, from the \textsc{BayeSN} fit to the $gri$ light curves. The shaded regions represent $1\sigma$ uncertainties on the mass step estimates. The results in red assume a single $R_{V} = 2.61$ for all SNe~Ia, while the results in blue assume $R_{V, \text{LM}} = 3.14$ for SNe~Ia in low mass host galaxies and $R_{V, \text{LM}} = 2.39$ for SNe~Ia in high mass host galaxies. The results are shown for a mass step location of $10^{10} \mathcal{M}_{\odot}$.}
    \label{fig:mass-step-by-wavelength}
\end{figure}

We visualise the inferred mass step as a function of wavelength for both the single and split $R_{V}$ cases in Figure \ref{fig:mass-step-by-wavelength}. This figure illustrates the non-zero mass step that persists across wavelength regime, regardless of assumptions about $R_{V}$, which suggests that dust is not the dominant driver of the mass step. The reduced mass step estimated assuming different values of $R_{V}$ for SNe~Ia in high mass vs low mass host galaxies relative to that estimated assuming a single $R_{V}$ for all objects, indicates that dust may play some role in explaining the mass step.

At present, the conclusions that can be drawn from this study are limited by the statistical power of the sample. In the future, larger samples of SNe~Ia that are well-observed in the $z$ band will enable better constraints on the mass step in the $z$ band. These observational results should be compared to robust forward model simulations of a dust-driven mass step and of a mass step driven by intrinsic differences among SNe~Ia to quantify the contribution of dust to the mass step. How to simulate SNe~Ia light curves under different assumption about the driver(s) of the mass step is a non-trivial question that is left to future work.

%% file: table3.tex
\begin{table*}
    \centering
    \caption{The mass step parameter estimates by subset of data used in the \textsc{BayeSN} fit for the ``Single $R_{V}$'' analysis variant, where it is assumed that $R_{V} = 2.61$ for all SNe~Ia. The individual-band distances that have been corrected for host galaxy dust extinction assuming $A_{V}$ from the $griz$ light curve fit. The mass step uncertainties include an additional uncertainty to account for uncertainty on the host galaxy mass estimates, though this term is sub-dominant compared to the statistical uncertainty. For the step location of $\mathcal{M}_{\text{step}} = 10^{10} \, \mathcal{M}_{\odot}$, there are 35 Foundation SNe~Ia and 14 YSE SNe~Ia in the low mass bin, and 70 Foundation SNe~Ia and 31 YSE SNe~Ia in the high mass bin. For the step location of $\mathcal{M}_{\text{step}} = 10^{\medianhostgalmass} \, \mathcal{M}_{\odot}$, there are 52 Foundation SNe~Ia and 23 YSE SNe~Ia in the low mass bin, and 53 Foundation SNe~Ia and 22 YSE SNe~Ia in the high mass bin.}
    \begin{tabular}{c|c|c|c|c|c|c|c}
    Fit & Sample & \multicolumn{2}{c}{$\sigma_{\text{res, LM}}$ [mag]} & \multicolumn{2}{c}{$\sigma_{\text{res, HM}}$ [mag]} & $\gamma$ [mag] & Significance \\
    \hline
    &  & Foundation & YSE & Foundation & YSE &  &  \\
    \hline
    \hline
    $\mathcal{M}_{\text{step}} = 10^{10} \, \mathcal{M}_{\odot}$ &  &  &  &  &  &  &  \\
    $\mu_{griz}$ & Foundation+YSE & $0.138 \pm 0.022$ & $0.086 \pm 0.048$ & $0.103 \pm 0.013$ & $0.217 \pm 0.036$ & $-0.126 \pm 0.027$ & 4.58 \\
     & Foundation & $0.138 \pm 0.021$ & -- & $0.103 \pm 0.013$ & -- & $-0.131 \pm 0.031$ & 4.16 \\
     & YSE & -- & $0.094 \pm 0.052$ & -- & $0.215 \pm 0.035$ & $-0.079 \pm 0.073$ & 1.08 \\
    $\mu_{gri}$ & Foundation+YSE & $0.138 \pm 0.023$ & $0.080 \pm 0.047$ & $0.107 \pm 0.014$ & $0.223 \pm 0.038$ & $-0.117 \pm 0.029$ & 4.06 \\
     & Foundation & $0.137 \pm 0.023$ & -- & $0.107 \pm 0.014$ & -- & $-0.131 \pm 0.032$ & 4.04 \\
     & YSE & -- & $0.090 \pm 0.051$ & -- & $0.220 \pm 0.038$ & $-0.046 \pm 0.074$ & 0.63 \\
    $\mu_{g}$ & Foundation+YSE & $0.159 \pm 0.029$ & $0.141 \pm 0.072$ & $0.117 \pm 0.019$ & $0.208 \pm 0.046$ & $-0.099 \pm 0.037$ & 2.72 \\
     & Foundation & $0.159 \pm 0.028$ & -- & $0.116 \pm 0.019$ & -- & $-0.124 \pm 0.038$ & 3.21 \\
     & YSE & -- & $0.145 \pm 0.079$ & -- & $0.192 \pm 0.046$ & $0.034 \pm 0.095$ & 0.36 \\
    $\mu_{r}$ & Foundation+YSE & $0.124 \pm 0.022$ & $0.138 \pm 0.065$ & $0.131 \pm 0.016$ & $0.233 \pm 0.041$ & $-0.102 \pm 0.031$ & 3.33 \\
     & Foundation & $0.124 \pm 0.022$ & -- & $0.131 \pm 0.016$ & -- & $-0.120 \pm 0.033$ & 3.63 \\
     & YSE & -- & $0.152 \pm 0.074$ & -- & $0.217 \pm 0.039$ & $0.009 \pm 0.085$ & 0.11 \\
    $\mu_{i}$ & Foundation+YSE & $0.150 \pm 0.023$ & $0.073 \pm 0.045$ & $0.102 \pm 0.013$ & $0.233 \pm 0.039$ & $-0.113 \pm 0.028$ & 4.00 \\
     & Foundation & $0.150 \pm 0.022$ & -- & $0.101 \pm 0.013$ & -- & $-0.125 \pm 0.032$ & 3.85 \\
     & YSE & -- & $0.079 \pm 0.050$ & -- & $0.222 \pm 0.038$ & $-0.036 \pm 0.071$ & 0.50 \\
    $\mu_{z}$ & Foundation+YSE & $0.147 \pm 0.024$ & $0.113 \pm 0.061$ & $0.115 \pm 0.015$ & $0.275 \pm 0.047$ & $-0.105 \pm 0.031$ & 3.37 \\
     & Foundation & $0.147 \pm 0.025$ & -- & $0.115 \pm 0.015$ & -- & $-0.118 \pm 0.034$ & 3.45 \\
     & YSE & -- & $0.129 \pm 0.068$ & -- & $0.249 \pm 0.041$ & $0.026 \pm 0.095$ & 0.27 \\
    \hline
    $\mathcal{M}_{\text{step}} = 10^{\medianhostgalmass} \, \mathcal{M}_{\odot}$ &  &  &  &  &  &  &  \\
    $\mu_{griz}$ & Foundation+YSE & $0.135 \pm 0.017$ & $0.191 \pm 0.043$ & $0.115 \pm 0.016$ & $0.196 \pm 0.041$ & $-0.082 \pm 0.027$ & 3.04 \\
     & Foundation & $0.134 \pm 0.017$ & -- & $0.115 \pm 0.016$ & -- & $-0.080 \pm 0.029$ & 2.73 \\
     & YSE & -- & $0.191 \pm 0.045$ & -- & $0.197 \pm 0.039$ & $-0.101 \pm 0.081$ & 1.25 \\
    $\mu_{gri}$ & Foundation+YSE & $0.136 \pm 0.018$ & $0.172 \pm 0.043$ & $0.120 \pm 0.017$ & $0.211 \pm 0.045$ & $-0.079 \pm 0.028$ & 2.81 \\
     & Foundation & $0.134 \pm 0.018$ & -- & $0.121 \pm 0.018$ & -- & $-0.078 \pm 0.031$ & 2.51 \\
     & YSE & -- & $0.174 \pm 0.047$ & -- & $0.216 \pm 0.045$ & $-0.087 \pm 0.078$ & 1.13 \\
    $\mu_{g}$ & Foundation+YSE & $0.143 \pm 0.022$ & $0.137 \pm 0.049$ & $0.136 \pm 0.023$ & $0.223 \pm 0.059$ & $-0.065 \pm 0.033$ & 1.97 \\
     & Foundation & $0.142 \pm 0.022$ & -- & $0.136 \pm 0.022$ & -- & $-0.073 \pm 0.036$ & 2.06 \\
     & YSE & -- & $0.145 \pm 0.049$ & -- & $0.214 \pm 0.056$ & $-0.014 \pm 0.081$ & 0.18 \\
    $\mu_{r}$ & Foundation+YSE & $0.122 \pm 0.018$ & $0.180 \pm 0.046$ & $0.147 \pm 0.019$ & $0.237 \pm 0.051$ & $-0.065 \pm 0.030$ & 2.14 \\
     & Foundation & $0.121 \pm 0.018$ & -- & $0.145 \pm 0.018$ & -- & $-0.076 \pm 0.032$ & 2.38 \\
     & YSE & -- & $0.185 \pm 0.047$ & -- & $0.222 \pm 0.044$ & $0.001 \pm 0.080$ & 0.01 \\
    $\mu_{i}$ & Foundation+YSE & $0.144 \pm 0.018$ & $0.171 \pm 0.042$ & $0.111 \pm 0.016$ & $0.226 \pm 0.047$ & $-0.080 \pm 0.028$ & 2.89 \\
     & Foundation & $0.143 \pm 0.019$ & -- & $0.110 \pm 0.015$ & -- & $-0.076 \pm 0.030$ & 2.56 \\
     & YSE & -- & $0.170 \pm 0.043$ & -- & $0.228 \pm 0.048$ & $-0.076 \pm 0.078$ & 0.97 \\
    $\mu_{z}$ & Foundation+YSE & $0.137 \pm 0.019$ & $0.195 \pm 0.050$ & $0.128 \pm 0.018$ & $0.275 \pm 0.057$ & $-0.074 \pm 0.030$ & 2.44 \\
     & Foundation & $0.136 \pm 0.019$ & -- & $0.128 \pm 0.018$ & -- & $-0.077 \pm 0.032$ & 2.41 \\
     & YSE & -- & $0.189 \pm 0.049$ & -- & $0.259 \pm 0.058$ & $-0.028 \pm 0.094$ & 0.29 \\
    \hline
    \hline
    \end{tabular}
    \label{tab:hubble-table-singlerv}
\end{table*}

%% file: table4.tex
\begin{table*}
    \centering
    \caption{The mass step parameter estimates by subset of data used in the \textsc{BayeSN} fit for the ``Split $R_{V}$'' analysis variant, where host galaxy dust extinction has been corrected for assuming $R_{V, \text{LM}} = 3.14$ for SNe~Ia in low mass host galaxies and $R_{V, \text{HM}} = 2.39$ for SNe~Ia in high mass host galaxies. Again, the single-band distances that have been corrected for host galaxy dust extinction assuming $A_{V}$ from the $griz$ split $R_{V}$ light curve fit, and the mass step uncertainties include an additional uncertainty to account for uncertainty on the host galaxy mass estimates.}
    \begin{tabular}{c|c|c|c|c|c|c|c}
    Fit & Sample & \multicolumn{2}{c}{$\sigma_{\text{res, LM}}$ [mag]} & \multicolumn{2}{c}{$\sigma_{\text{res, HM}}$ [mag]} & $\gamma$ [mag] & Significance \\
    \hline
    &  & Foundation & YSE & Foundation & YSE &  &  \\
    \hline
    \hline
    $\mathcal{M}_{\text{step}} = 10^{10} \, \mathcal{M}_{\odot}$ &  &  &  &  &  &  &  \\
    $\mu_{griz}$ & Foundation+YSE & $0.132 \pm 0.022$ & $0.092 \pm 0.050$ & $0.104 \pm 0.013$ & $0.217 \pm 0.036$ & $-0.102 \pm 0.027$ & 3.78 \\
     & Foundation & $0.132 \pm 0.022$ & -- & $0.105 \pm 0.013$ & -- & $-0.106 \pm 0.031$ & 3.41 \\
     & YSE & -- & $0.105 \pm 0.058$ & -- & $0.211 \pm 0.036$ & $-0.055 \pm 0.074$ & 0.74 \\
    $\mu_{gri}$ & Foundation+YSE & $0.138 \pm 0.023$ & $0.078 \pm 0.048$ & $0.109 \pm 0.014$ & $0.220 \pm 0.038$ & $-0.088 \pm 0.029$ & 3.03 \\
     & Foundation & $0.137 \pm 0.024$ & -- & $0.109 \pm 0.014$ & -- & $-0.100 \pm 0.033$ & 3.06 \\
     & YSE & -- & $0.083 \pm 0.051$ & -- & $0.215 \pm 0.037$ & $-0.015 \pm 0.073$ & 0.20 \\
    $\mu_{g}$ & Foundation+YSE & $0.158 \pm 0.029$ & $0.139 \pm 0.074$ & $0.120 \pm 0.019$ & $0.214 \pm 0.047$ & $-0.079 \pm 0.036$ & 2.16 \\
     & Foundation & $0.158 \pm 0.030$ & -- & $0.119 \pm 0.018$ & -- & $-0.107 \pm 0.038$ & 2.80 \\
     & YSE & -- & $0.141 \pm 0.075$ & -- & $0.196 \pm 0.045$ & $0.054 \pm 0.093$ & 0.58 \\
    $\mu_{r}$ & Foundation+YSE & $0.124 \pm 0.022$ & $0.138 \pm 0.066$ & $0.134 \pm 0.016$ & $0.235 \pm 0.041$ & $-0.072 \pm 0.031$ & 2.35 \\
     & Foundation & $0.124 \pm 0.021$ & -- & $0.134 \pm 0.016$ & -- & $-0.088 \pm 0.033$ & 2.71 \\
     & YSE & -- & $0.156 \pm 0.070$ & -- & $0.217 \pm 0.040$ & $0.046 \pm 0.087$ & 0.53 \\
    $\mu_{i}$ & Foundation+YSE & $0.151 \pm 0.023$ & $0.072 \pm 0.045$ & $0.103 \pm 0.013$ & $0.233 \pm 0.039$ & $-0.089 \pm 0.028$ & 3.16 \\
     & Foundation & $0.152 \pm 0.023$ & -- & $0.103 \pm 0.013$ & -- & $-0.100 \pm 0.033$ & 3.04 \\
     & YSE & -- & $0.080 \pm 0.050$ & -- & $0.224 \pm 0.036$ & $-0.006 \pm 0.073$ & 0.08 \\
    $\mu_{z}$ & Foundation+YSE & $0.148 \pm 0.024$ & $0.113 \pm 0.061$ & $0.116 \pm 0.015$ & $0.276 \pm 0.048$ & $-0.086 \pm 0.032$ & 2.73 \\
     & Foundation & $0.148 \pm 0.024$ & -- & $0.115 \pm 0.015$ & -- & $-0.099 \pm 0.033$ & 2.96 \\
     & YSE & -- & $0.124 \pm 0.069$ & -- & $0.246 \pm 0.041$ & $0.045 \pm 0.093$ & 0.49 \\
    \hline
    $\mathcal{M}_{\text{step}} = 10^{\medianhostgalmass} \, \mathcal{M}_{\odot}$ &  &  &  &  &  &  &  \\
    $\mu_{griz}$ & Foundation+YSE & $0.132 \pm 0.018$ & $0.202 \pm 0.046$ & $0.116 \pm 0.016$ & $0.201 \pm 0.041$ & $-0.055 \pm 0.027$ & 2.01 \\
     & Foundation & $0.132 \pm 0.017$ & -- & $0.116 \pm 0.015$ & -- & $-0.054 \pm 0.030$ & 1.79 \\
     & YSE & -- & $0.202 \pm 0.048$ & -- & $0.200 \pm 0.041$ & $-0.060 \pm 0.082$ & 0.74 \\
    $\mu_{gri}$ & Foundation+YSE & $0.141 \pm 0.019$ & $0.184 \pm 0.046$ & $0.120 \pm 0.017$ & $0.213 \pm 0.045$ & $-0.044 \pm 0.029$ & 1.54 \\
     & Foundation & $0.140 \pm 0.018$ & -- & $0.120 \pm 0.017$ & -- & $-0.044 \pm 0.032$ & 1.40 \\
     & YSE & -- & $0.185 \pm 0.047$ & -- & $0.215 \pm 0.044$ & $-0.047 \pm 0.078$ & 0.61 \\
    $\mu_{g}$ & Foundation+YSE & $0.143 \pm 0.022$ & $0.142 \pm 0.050$ & $0.138 \pm 0.023$ & $0.233 \pm 0.059$ & $-0.039 \pm 0.033$ & 1.18 \\
     & Foundation & $0.142 \pm 0.023$ & -- & $0.137 \pm 0.021$ & -- & $-0.049 \pm 0.037$ & 1.32 \\
     & YSE & -- & $0.148 \pm 0.051$ & -- & $0.221 \pm 0.058$ & $0.020 \pm 0.080$ & 0.25 \\
    $\mu_{r}$ & Foundation+YSE & $0.124 \pm 0.018$ & $0.196 \pm 0.048$ & $0.148 \pm 0.019$ & $0.243 \pm 0.052$ & $-0.030 \pm 0.031$ & 0.99 \\
     & Foundation & $0.124 \pm 0.019$ & -- & $0.147 \pm 0.020$ & -- & $-0.041 \pm 0.033$ & 1.22 \\
     & YSE & -- & $0.200 \pm 0.049$ & -- & $0.231 \pm 0.049$ & $0.044 \pm 0.083$ & 0.53 \\
    $\mu_{i}$ & Foundation+YSE & $0.147 \pm 0.018$ & $0.185 \pm 0.044$ & $0.111 \pm 0.016$ & $0.228 \pm 0.048$ & $-0.050 \pm 0.028$ & 1.81 \\
     & Foundation & $0.147 \pm 0.019$ & -- & $0.112 \pm 0.016$ & -- & $-0.050 \pm 0.031$ & 1.59 \\
     & YSE & -- & $0.186 \pm 0.044$ & -- & $0.223 \pm 0.044$ & $-0.034 \pm 0.080$ & 0.43 \\
    $\mu_{z}$ & Foundation+YSE & $0.139 \pm 0.019$ & $0.201 \pm 0.050$ & $0.128 \pm 0.018$ & $0.278 \pm 0.058$ & $-0.052 \pm 0.030$ & 1.71 \\
     & Foundation & $0.138 \pm 0.019$ & -- & $0.128 \pm 0.018$ & -- & $-0.055 \pm 0.033$ & 1.69 \\
     & YSE & -- & $0.199 \pm 0.050$ & -- & $0.260 \pm 0.055$ & $-0.000 \pm 0.094$ & 0.00 \\
    \hline
    \hline
    \end{tabular}
    \label{tab:hubble-table-splitrv}
\end{table*}

%% file: 05_discussion.tex
We now discuss the results of this analysis in relation to previous work investigating properties of SNe~Ia across wavelength space. In \S\ref{sec:discussion-stretchlum}, we discuss the pre-correction residual scatter in the intrinsic magnitudes and their dependence on shape. In \S\ref{sec:discussion-HD}, we compare the total scatter in the Hubble diagrams across wavelength space. Finally in \S\ref{sec:discussion-massstep}, we discuss previous estimates of the mass step in comparison to the mass step inferred in this work.

\subsection{Standardisation of SNe~Ia Peak $z$-band Magnitudes}
\label{sec:discussion-stretchlum}
We find that the residual scatter in the peak extinguished $z$-band magnitudes is smaller than in the $g$- and $r$-bands and larger than in the $i$-band. However, studies of the standardisation of SNe~Ia in the NIR consistently find lower levels of residual scatter in the peak extinguished absolute magnitude in the NIR $YJH$ filters than this work reports in the $z$-band, regardless of differences in sample and methods \citep{Dhawan_2018, Stanishev_2018, Avelino_2019}. While the $i$- and $z$-bands therefore do not appear to be more standard pre-correction for dust and light curve shape/colour effects compared to the NIR, they do represent a improvement from the optical $g$- and $r$-bands. 

Additionally, we find a significant correlation between $M_{z, \text{int}}$ and $\theta_{gri}$, with a fitted linear model having a slope of $a = \zStretchLumSlope \pm \zStretchLumSlopeErr$. Because different methods for light curve fitting use different definitions of a shape parameter, and this analysis is the first to use \textsc{BayeSN} to estimate the shape-luminosity relation, we can only compare qualitatively to the results from previous analyses. There is broad support for relatively weak shape-luminosity relations in the NIR (YJH) compared to in the optical \citep{Kattner_2012, Stanishev_2018, Burns_2018, Uddin_2023, Peterson_2024}. This work affirms the presence of a shape-luminosity relation in the $z$-band, though understanding the strength of this correlation relative to optical and redder NIR filters will require data from SNe~Ia observed from the optical through the NIR, including the $z$-band.

\subsection{The Hubble Diagram}
\label{sec:discussion-HD}
We find that the $z$-band Hubble diagram has a total scatter of $\zRMS$ mag. Comparing to the total scatter in the other individual-band Hubble diagrams, we find that SNe~Ia in the $z$ band are not relatively better standard candles than in the $i$ band. There are hints that SNe~Ia in the $z$ band may be relatively better standard candles than in the $g$ and $r$ bands, but this result cannot be decisively claimed with the sample in this work. Notably, trends in the total and residual scatter across wavelength regimes differ for the Foundation and YSE subsets of the data, which we attribute to differences in the two surveys (e.g., being targeted vs untargeted).

We also find that adding the $z$-band data to construct a Hubble diagram from the full $griz$ data results in an decrease of 0.005 mag in the RMS compared to a Hubble diagram constructed from $gri$ data only. This result is broadly consistent with previous work demonstrating the benefit of combining optical and NIR data to reduce scatter in the Hubble diagram \citep{Avelino_2019, Thorp_2021, Mandel_2022, Dhawan_2023, Uddin_2023}.

Because of differences in samples and methods used to standardise the SNe~Ia across analyses, it is difficult to compare quantitatively for a determination of how standard SNe~Ia in the $z$-band are relative to in the NIR. For such an analysis, a large uniform sample of SNe~Ia observed from the optical to the NIR, including the $z$ band, is necessary. The upcoming Rubin-LSST, expected to begin survey operations in October 2025, will provide the necessary wavelength coverage and statistical leverage to draw stronger conclusions regarding the relative standardisation of SNe~Ia across wavelength \citep{Ivezic_2019}. Still, this work provides further evidence for the benefit of having data in redder filters (from the $z$-band to the NIR) in reducing the scatter in the Hubble residuals, as suggested by previous studies.

\subsection{The Mass Step}
\label{sec:discussion-massstep}
We constrain the mass step to be $\gamma_{griz} = \grizMassStepTen \pm \grizMassStepErrTen$ mag for a mass step location of $10^{10} \mathcal{M}_{\odot}$ in the fiducial analysis. Across choices of mass step location and differential treatments of dust for high and low mass hosts, this analysis finds evidence of a consistent mass step from the optical to the $z$-band at 1-4$\sigma$ significance. Previous analyses of the Foundation sample have yielded mass steps of similar size and significance as is reported in this analysis \citep{Jones_2019, Thorp_2021}.

Recently, DES, a five-year survey which observed the sky in the $griz$ filters with the Dark Energy Camera (DECam), produced a sample 1499 photometrically classified SNe~Ia \citep{Vincenzi_2024}. In their cosmological analysis, they correct for a mass step of size $-0.046 \pm 0.009$ at $10^{10} \mathcal{M}_{\odot}$. Also, \citet{Ginolin_2024} recently analysed the mass step in a sample of 927 spectroscopically-classified SNe~Ia observed in $gri$ from the Zwicky Transient Facility \citep[ZTF;][]{Bellm_2019} Data Release 2 \citep[DR2;][]{Rigault_2025}. They find a much larger mass step of $-0.162 \pm 0.020$ at $10^{10} \mathcal{M}_{\odot}$ in a similar wavelength regime, though for a much lower redshift sample.

We largely attribute the difference seen across these analyses to the different samples. The selection function for each of these surveys will be slightly different based on whether the sample is targeted (i.e. Foundation, DES) vs. untargeted (i.e. YSE, ZTF), or spectroscopically (i.e. Foundation, YSE, ZTF) vs. photometrically classified (i.e. DES). These differences may affect the samples enough to result in the difference in the reported mass steps. In addition, \citet{Vincenzi_2024} and \citet{Ginolin_2024} use \textsc{SALT3} and \textsc{SALT2.4} models, respectively, to fit for the distance moduli \citep{Kenworthy_2021, Taylor_2023}, while this analysis uses \textsc{BayeSN}. These differences should be investigated in more detail to determine how they impact mass step inference \citep[e.g.,][ for a comparison of \textsc{SALT} and \textsc{BayeSN}]{Grayling_Popovic_2024}. Furthermore, \citet{Murakami_2025} recently raised that differences in analyses could lead to different results even in the same sample, so comparisons across works must be carefully considered.

In the NIR $Y$-, $J$-, and $H$-bands, mass steps of $-0.02$ to $-0.15$ mag have been reported for mass step locations of $10^{10-10.6} \mathcal{M}_{\odot}$ \citep{Uddin_2020, Ponder_2021, Johansson_2021, TM_2022, Jones_2022, Uddin_2023, Peterson_2024}. The approaches for estimating a NIR-only mass step have varied in their choices (e.g., whether they correct for a shape-luminosity relation, colour effects, or host galaxy dust; whether NIR filters are fit individually or jointly; choice of light curve fitting method).

Despite their differences, the majority of these works agree that there is evidence for a non-zero NIR mass step which is of a similar size as the optical mass step \citep{Uddin_2020, Ponder_2021, Jones_2022, TM_2022}. Based on this result, differences in dust properties in high- and low-mass host galaxies are not likely to be the physical cause for the mass step, as had been suggested by \citet{Brout_Scolnic_2021}. The low statistical significance of the step size reported in these studies limits the strength of this claim. The NIR mass step found in these analyses are largely consistent within $1\sigma$ with the mass step reported in this work and with the size of the optical mass step found in each of their samples.

However, several works report mass steps which are consistent with zero in the NIR \citep{Johansson_2021, Uddin_2023, Peterson_2024}. \citet{Johansson_2021} and \citet{Uddin_2023} report evidence for a mass step that is non-zero in the optical, but which is consistent with zero and the optical mass step in the NIR. \citet{Peterson_2024} report a mass step that is consistent with zero in both the optical and NIR wavelength regimes. The size of the mass steps reported in these works are consistent with the size of the optical and $z$-band mass steps reported in this work despite differences in sample, light curve fitting models, and methods for correcting for dust.

As in this analysis of the optical and $z$ band mass steps, these previous studies of the mass step from the optical to the NIR largely do not constrain a change in the mass step as a function of wavelength, regardless of dust treatment. This result disfavours the hypothesis that dust explains the observed mass step in full given the current statistical precision. Though dust may play some role in explaining the mass step, as indicated by the lessened optical and $z$-band mass step for the split $R_{V}$ analysis in this work, dust alone does not appear to explain the mass step. 

As previously mentioned, these results should be compared to those from robust forward simulations of the mass step under different assumptions about its origin. Recently, \citet{Peterson_2024} used the \texttt{SNANA} package for cosmological analyses with SNe~Ia \citep{Kessler_2009_SNANA, Kessler_2019} to simulate the DEHVILS survey used in their analysis for two models of the mass step. Firstly, they assumed a dust-based mass step, á la \citet{Brout_Scolnic_2021}, and secondly, a "achromatic" mass step, based on artificial variations in the SN~Ia SED that result in a wavelength-independent scatter. \citet{Peterson_2024} found their sample did not have sufficient precision to affirm or rule out dust as the origin of the mass step. The creation of realistic forward simulations and the statistical precision allowed by a sample of SNe~Ia should be considered in such tests in the future.

The main limitation of current studies, though, is the availability of data in redder wavelength regimes. Studies of the mass step with some of the largest uniform samples of SNe~Ia in the NIR have not been able to find a significant change in the size of the mass step from the optical to redder wavelength regimes. While this result provides some support for an intrinsic explanation of the mass step, it is possible that different $R_{V}$ values in high and low mass hosts produce a non-zero mass step in the NIR \citep{Uddin_2023, Peterson_2024}. Still, the NIR mass step would be lessened compared to the optical mass step in this case. Larger samples of SNe~Ia observed across optical and NIR will be necessary to constrain small deviations of the mass step across wavelength space to definitively confirm or rule out a dust-based explanation for the mass step.

With larger samples of SNe~Ia observed across a broad wavelength regime, the assumption of a mass ``step'' rather than another model for the Hubble residual dependence on host galaxy mass (e.g., a linear trend) should also be challenged. In this work, we consider only a step function because of the clear predictions made by \citet{Brout_Scolnic_2021} regarding the behaviour of a mass step as a function of wavelength, assuming the step is caused by differences in host galaxy dust properties. An alternative description of the relationship between supernova magnitude and host mass may better describe the physical process producing this observed residual dependence.

We proceed with a step in this work because a two-population model might suggest a step if mass traces the two populations cleanly \citep{Briday_2022}. That said, a more complex model may be necessary if global host mass does not cleanly trace the two populations, which may arise if e.g., mass is determined to be a poor tracer or there exist more than two populations within the data. Indeed, \citep{Murakami_2025} recently presented a method for modelling a continuous evolution in the Hubble residuals with host properties, which they applied to the ZTF DR2 sample. In the future, more data across the optical and NIR will be necessary \citep{Phillips_2019} to determine if a different model is preferred over the fiducial step function model. Rubin-LSST is expected to provide an ideal sample for such further tests in its ten-year lifetime.

Finally, we note that global host galaxy mass is not the strongest tracer of the mass step \citep{Briday_2022, Wiseman_2023}. Indeed, other tracers -- such as local mass or global/local star formation rate -- may provide additional insight into how best to standardise SNe~Ia \citep{Sullivan_2003, Gallagher_2008, Childress_2013_metallicity, Rigault_2013, MorenoRaya_2016, Rose_2019, Hakobyan_2020, Rose_2021, Kelsey_2021, MillanIrigoyen_2022, Briday_2022, Wiseman_2023, Kelsey_2023}.

%% file: 06_conclusion.tex
In this paper, we explore the cosmological utility of SNe~Ia the $z$-band. Observations in the $z$-band benefit from a reduced sensitivity to dust compared to the optical, yet can be obtained to greater depth than NIR filters using CCDs. Therefore, the $z$-band sits at a sweet spot between the optical and NIR. Despite this, it has yet to be extensively studied. We present a fully Bayesian framework for analysing the properties of SNe~Ia in the $z$-band, which can be applied to any cosmological analysis of SNe~Ia and may be particularly useful to future studies of SNe~Ia in an individual filter\footnote{The code for this model can be found on GitHub at \url{https://github.com/erinhay/snia-zband}.}.

We find that both the peak extinguished magnitudes and intrinsic magnitudes in the $z$-band are relatively better standard compared to the $g$-band and $r$-band, and exhibit a similar level of variance compared to the $i$-band. The residual scatter in the peak $z$-band extinguished magnitudes is $\sigma_{\text{res, ext}}(z) = \zExtResScat \pm \zExtResScatErr$ mag, compared to $\sigma_{\text{res, ext}}(g) = \gExtResScat \pm \gExtResScatErr$ mag, $\sigma_{\text{res, ext}}(r) = \rExtResScat \pm \rExtResScatErr$ mag, and $\sigma_{\text{res, ext}}(i) = \iExtResScat \pm \iExtResScatErr$ mag for the $g$-band, $r$-band and $i$-band, respectively. We also find that the $z$-band magnitudes at peak that have been corrected for extinction from dust in the host galaxy are correlated with the inferred light curve shape parameter. The linear shape-luminosity relation fit has a slope of $\zStretchLumSlope \pm \zStretchLumSlopeErr$ with a residual scatter around the relation of $\zStretchLumScat \pm \zStretchLumScatErr$ mag. The $z$-band peak intrinsic magnitudes do, therefore, exhibit some dependence on the shape parameter. 

We then construct the first $z$-band only Hubble diagram and find a total scatter of RMS $= \zRMS$. In addition, the Hubble diagram shows less total scatter with optical and $z$-band data than when using only optical data. When combined with host galaxy mass information, we are able to constrain the size of the mass step in the $z$-band to be $\gamma_{z} = \zMassStepTen \pm \zMassStepErrTen$ mag for a step location of $\mathcal{M}_{\text{step}} = 10^{10} \mathcal{M}_{\odot}$ when assuming $R_{V} = 2.61$ for all SNe~Ia. This mass step is consistent with the mass steps in each of the $g$, $r$, and $i$ bands. When assuming different $R_{V}$ values for SNe~Ia in high mass and low mass host galaxies, we estimate $\gamma_{z} = \zSplitMassStepTen \pm \zSplitMassStepErrTen$ for the same choice of step location. While the split $R_{V}$ case leads to a reduced mass step compared to the single $R_{V}$ case, the optical mass step is consistent with the $z$-band mass step to within $1\sigma$ regardless of assumptions about $R_{V}$. Therefore, this work suggests dust alone cannot explain the mass step.

While SNe~Ia in the $z$-band do not appear to be standard candles, based on their dependence on the shape parameter, this analysis demonstrates that the $z$-band wavelength regime exhibits complementary properties to the optical. Observations in the $z$-band appear to improve distances constraints to SNe~Ia, as the $griz$ Hubble diagram exhibits a smaller amount of total scatter than the optical only $gri$ Hubble diagram. Still, SNe~Ia in the NIR do appear to be relatively better standard candles than SNe~Ia in the $z$-band, as demonstrated by a higher total scatter in the $z$-band Hubble residuals compared to that in NIR Hubble residuals from other works. In the future, observations which span the optical, the $z$-band, and the NIR for the same sample of SNe~Ia will be necessary to understand the relative standardisation properties of SNe~Ia across these wavelength regimes.

As YSE is an ongoing survey, it will continue to discover and make follow-up observations of SNe~Ia, so the sample of SNe~Ia observed in the $z$-band is growing by the day. A larger sample size will greatly improve the precision of studies of the standardisation properties of SNe~Ia in the $z$-band and potentially allow for a more definitive determination of the contribution of dust to the post-standardisation dependence of the Hubble residuals on host galaxy mass. Furthermore, there is a sample of over 1,000 photometrically confirmed SNe~Ia included in DR1 \citep{Aleo_2023}. Although subject to contamination from non-Ias, the additional objects may prove beneficial to a study of the mass step in the $z$-band regardless. We leave the analysis of this sample and future data releases from YSE to future work.

More broadly, the $z$-band is increasingly important to understand for the modelling of the SNe~Ia that will be observed by upcoming cosmological surveys. For example, Rubin-LSST will observe SNe~Ia in the rest-frame $z$-band out to $z \sim 0.1$ \citep{Ivezic_2019}. This low-redshift sample of SNe~Ia will serve as the anchor for crucial cosmological analyses in the future. Additionally, the Roman Space Telescope will observe SNe~Ia out to $z \sim 1.8$ \citep{Hounsell_2018}, enabling precision studies of SNe~Ia in redder wavelength regimes. Therefore, it is important that our models for the light curves, such as \textsc{BayeSN}, are trained on sufficient $z$-band data so that the behaviour of SNe~Ia in this wavelength regime is well-constrained to take full advantage of the sample of SNe~Ia from these upcoming instruments. Accurate and comprehensive models of SNe~Ia trained across a wide wavelength regime and large samples of SNe~Ia from next-generation cosmological surveys will enable estimates of the Hubble parameter and the dark energy equation of state to unprecedented precision in the next decade.

%% file: 07_appendixA.tex
In Table \ref{tab:full-table}, we present an abridged version of the data used in this analysis. The table includes metadata for each SN~Ia, including redshift and host galaxy mass, as well as fitted parameters, including the \textsc{BayeSN} parameters from each of the $griz$, $gri$, and individual-band fits and the peak magnitudes in each filter. The full table can be found online.

%% file: 08_appendixB.tex
Here, we elaborate on the motivation and justification for the $z$-band data requirement, as described in \S\ref{sec:data-cuts}. Because we fit the $z$-band only light curves, we need to ensure the SNe~Ia are well enough sampled in the $z$ band to obtain a reliable distance estimate. Therefore, we set up a simple test to determine the amount of $z$-band data necessary to get a reliable distance estimate, which we describe below.

We simulate 80-100 light curves from the \textsc{BayeSN} model with $N$ $z$-band observations, where $N \in {1, 2, ..., 7}$. The times of observations are randomly sampled from between -5 and 15 days. We then fit these 661 unique light curves with \textsc{BayeSN}. Binning by number of $z$-band observations, we examine how many $2\sigma$ and $3\sigma$ outliers there are in recovered distance modulus. Regardless of number of $z$-band observations, the percent of $2\sigma$ outliers lies between 5\% and 9\% and does not show trends with number of $z$-band observations. Similarly, the $3\sigma$ outliers is low (between 0-1.5\%) and consistent across number of observations in the light curve.

As the quantity of outliers is well calibrated across any number of observations in the light curve, we examine the distribution of uncertainty on the distance modulus for the 80-100 light curves per bin. The mean uncertainty consistently decreases from 0.22 for 1 $z$-band observation to below 0.18 for more than 5 $z$-band observations. From this metric, there is not an obvious choice of number of $z$-band observations to require, though the mean uncertainty drops below 0.2 for 3 $z$-band observations. The standard deviation of the distribution, however, drops from 0.026 for 2 $z$-band observations to 0.017 for 3 $z$-band observations -- the biggest reduction when moving from 1 $z$-band observation to 7 $z$-band observations. Therefore, we proceed with the requirement of 3 $z$-band observations.

Additionally, we test the results of this analysis when making a more strict cut on the amount of $z$-band data to confirm the robustness of the results. When we require 4 $z$-band observations, the sample is reduced to 124 SNe~Ia, with 25 from YSE and 99 from Foundation. We find that the conclusions of the analysis remain unchanged. In particular, we find the $griz$ mass step to be $-0.064 \pm 0.028$ mag and the $z$-band mass step to be $-0.061 \pm 0.031$ mag at $10^{10} M_{\odot}$. We do not include an uncertainty arising from host galaxy mass estimate uncertainties here, though we do not expect it to be significant given it is a sub-dominant source of uncertainty in the main analysis. These results are of reduced significance than the fiducial analysis, but broadly align with the range of mass step estimates inferred in this work. The conclusions of the analysis remain unchanged.

In the GitHub repository associated with this paper, at \url{https://github.com/erinhay/snia-zband}, we include example notebooks for running the analysis which allow the user to change the sample according to their own data cuts. To test additional iterations of this analysis, we encourage the use of this functionality.

%% file: 09_appendixC.tex
\subsection{Fixed Theta Fits}

Because we use only a subset of the data in the $z$-band only fit, the constraint on $\theta$ is not going to be the most accurate. Here, we explore the impact of fixing $\theta$ to the value determined from the fit to the optical $gri$ data, which we will denote $\theta_{\text{gri}}$, when fitting with $z$-band only data. We use $\theta$ from the $gri$ fit, rather than from the $griz$ fit, to avoid double-counting the $z$-band data. 

As expected, the $z$-band Hubble residuals from the fixed $\theta$ \textsc{BayeSN} fits show less total scatter than those from the fit where $\theta$ is allowed to vary freely. A reduction of the Hubble residuals RMS from $\zRMS$ to $0.190$ is seen overall, with the RMS for the Foundation subset being $0.156$ and for the YSE subset being $0.252$. We also see a more significant detection of the mass step in the $z$ band with the fixed $\theta$ fits -- at $\gamma_{z, \text{fixed} \; \theta} = -0.145 \pm 0.029$ mag for a step at $10^{10} \mathcal{M}_{\odot}$. We do not include uncertainty arising from host galaxy mass estimate uncertainties here, though we do not expect it to be significant. This result leaves the conclusions of this work unchanged.

\subsection{Peak Magnitudes from $griz$ Data}

Alternatively to retrieving the $z$-band peak apparent magnitude from the $z$-band \textsc{BayeSN} fit, we can extract the peak apparent magnitude in the $z$ band from the fit to the full $griz$ light curve, as described in \S\ref{sec:analysis-hubblediagram}. This approach relies more heavily on the assumed light curve model because the trained relation between the shape of the light curve in the $gri$ filters and the shape in the $z$ band will influence the resulting peak magnitudes. However, it allows us to leverage the multi-band data to get the best constraints on $\theta$ and $A_{V}$.

When using the $griz$ fit, the fitted mean peak intrinsic absolute magnitude in the $z$ band is found to be $M^{0}_{z, \text{int}} = -18.31 \pm 0.02$ mag, with a residual scatter of $\sigma_{\text{res, int}}(z) = 0.24 \pm 0.02$. Under this alternative peak $z$-band magnitude extraction routine, the results of the analysis remain consistent with the fiducial analysis (c.f. Table \ref{tab:peak-mags-table}).

%% file: tableA1.tex
\begin{landscape}
\begin{table}
    \centering
    \renewcommand\thetable{A1}
    \caption{The metadata and fitted light curve parameters for a subset of the SNe~Ia in the sample. All light curve parameters listed in the abridged table below are fit assuming $R_{V} = 2.61$. The full table with information for all SNe~Ia is available online.}
    \begin{tabular}{c|c|c|c|c|c|c|c|c|c|c|c}
        SNID & RA & DEC & Survey & $z_{\text{final}}$ & $\sigma_{z}$ & $\log_{10}(\mathcal{M}_{*}/\mathcal{M}_{\odot})$ & $\mu_{griz}$ & $A_{V, griz}$ & $\theta_{griz}$ & $T_{0, griz}$ & Reduced $\chi^{2}_{griz}$ \\
        \hline
        \hline
        2016W & 37.6654873 & 42.2359685 & Foundation & 0.019 & 0.00006700 & 11.008 & $34.477 \pm 0.047$ & $0.223 \pm 0.056$ & $1.275 \pm 0.144$ & $57417.927 \pm 0.198$ & 1.200 \\
        2016afk & 155.6245508 & 15.0966103 & Foundation & 0.048 & 0.00000670 & 9.782 & $36.603 \pm 0.063$ & $0.236 \pm 0.072$ & $-0.075 \pm 0.189$ & $57439.853 \pm 0.491$ & 1.136 \\
        2020adkc & 154.081739 & -1.516317 & YSE & 0.092 & 0.00021300 & 10.548 & $37.680 \pm 0.076$ & $0.115 \pm 0.076$ & $-0.886 \pm 0.238$ & $59218.175 \pm 0.569$ & 0.761 \\
        2020evu & 238.12045 & 24.494119 & YSE & 0.059 & 0.00000659 & 10.261 & $36.891 \pm 0.077$ & $0.450 \pm 0.076$ & $-0.646 \pm 0.183$ & $58941.498 \pm 0.389$ & 0.424 \\
        2020ewx & 229.327104 & 13.02275 & YSE & 0.061 & 0.01000000 & 9.556 & $37.053 \pm 0.059$ & $0.143 \pm 0.086$ & $-0.830 \pm 0.256$ & $58944.728 \pm 0.908$ & 0.331 \\
        2020fhs & 240.276175 & 19.450186 & YSE & 0.015 & 0.00000678 & 10.364 & $34.315 \pm 0.065$ & $0.510 \pm 0.076$ & $2.100 \pm 0.174$ & $58949.045 \pm 0.361$ & 0.214 \\
        2020ilb & 145.266392 & 34.73362 & YSE & 0.052 & 0.00001000 & 11.073 & $36.435 \pm 0.065$ & $0.557 \pm 0.081$ & $-0.115 \pm 0.357$ & $58980.837 \pm 0.563$ & 0.203 \\
        2020ioz & 195.94825 & 20.902494 & YSE & 0.061 & 0.01000000 & 9.474 & $37.361 \pm 0.045$ & $0.067 \pm 0.047$ & $-0.981 \pm 0.190$ & $58976.940 \pm 0.352$ & 0.575 \\
        2020rii & 249.016723 & 33.209832 & YSE & 0.086 & 0.00000905 & 10.591 & $37.687 \pm 0.061$ & $0.176 \pm 0.066$ & $0.574 \pm 0.258$ & $59077.994 \pm 0.345$ & 2.353 \\
        2020rmg & 17.280127 & -15.997169 & YSE & 0.038 & 0.00011300 & 11.214 & $36.133 \pm 0.062$ & $0.079 \pm 0.052$ & $-0.009 \pm 0.174$ & $59091.215 \pm 0.369$ & 1.140 \\
        \hline
    \end{tabular}
    \label{tab:full-table}
\end{table}

\begin{table}
    \centering
    \renewcommand\thetable{A1 (cont)}
    \caption{Table A1 continued.}
    \begin{tabular}{c|c|c|c|c|c|c|c|c|c|c|c|c|c|c}
        SNID & $\mu_{gri}$ & $A_{V, gri}$ & $\theta_{gri}$ & $T_{0, gri}$ & $\mu_{z}$ & $\theta_{z}$ & \multicolumn{4}{c}{Number of Obs.} & \multicolumn{4}{c}{Time of First Obs. [MJD]} \\
         &  &  &  &  &  &  & $g$ & $r$ & $i$ & $z$ & $g$ & $r$ & $i$ & $z$ \\
        \hline
        \hline
        2016W & $34.449 \pm 0.065$ & $0.240 \pm 0.069$ & $1.349 \pm 0.156$ & $57417.906 \pm 0.206$ & $34.623 \pm 0.065$ & $1.000 \pm 0.250$ & 7 & 7 & 7 & 7 & 57410.29 & 57410.29 & 57410.29 & 57410.29 \\
        2016afk & $36.554 \pm 0.079$ & $0.297 \pm 0.089$ & $-0.221 \pm 0.224$ & $57439.637 \pm 0.571$ & $36.779 \pm 0.060$ & $-0.169 \pm 0.328$ & 7 & 11 & 7 & 6 & 57437.40 & 57437.40 & 57437.40 & 57437.40 \\
        2020adkc & $37.677 \pm 0.079$ & $0.101 \pm 0.074$ & $-0.829 \pm 0.243$ & $59218.212 \pm 0.588$ & $37.753 \pm 0.129$ & $-0.630 \pm 0.789$ & 4 & 7 & 4 & 3 & 59230.60 & 59208.55 & 59212.64 & 59208.54 \\
        2020evu & $36.992 \pm 0.141$ & $0.367 \pm 0.122$ & $-0.704 \pm 0.212$ & $58941.395 \pm 0.382$ & $37.082 \pm 0.088$ & $-0.448 \pm 0.463$ & 8 & 7 & 4 & 6 & 58931.52 & 58931.52 & 58958.59 & 58941.54 \\
        2020ewx & $36.998 \pm 0.079$ & $0.207 \pm 0.113$ & $-0.867 \pm 0.276$ & $58944.965 \pm 0.917$ & $37.135 \pm 0.136$ & $-0.086 \pm 0.927$ & 5 & 5 & 4 & 3 & 58955.46 & 58945.50 & 58942.49 & 58951.45 \\
        2020fhs & $34.367 \pm 0.098$ & $0.401 \pm 0.113$ & $2.326 \pm 0.204$ & $58949.012 \pm 0.378$ & $34.766 \pm 0.077$ & $1.412 \pm 0.285$ & 5 & 4 & 4 & 4 & 58959.59 & 58947.61 & 58959.59 & 58947.62 \\
        2020ilb & $36.405 \pm 0.119$ & $0.599 \pm 0.122$ & $-0.154 \pm 0.370$ & $58980.760 \pm 0.547$ & $36.689 \pm 0.141$ & $-0.013 \pm 0.935$ & 3 & 3 & 3 & 3 & 58985.26 & 58973.27 & 58973.27 & 58974.31 \\
        2020ioz & $37.330 \pm 0.063$ & $0.066 \pm 0.043$ & $-0.842 \pm 0.217$ & $58976.953 \pm 0.357$ & $37.463 \pm 0.068$ & $-1.239 \pm 0.421$ & 7 & 5 & 6 & 4 & 58971.29 & 58981.41 & 58969.32 & 58971.29 \\
        \hline
    \end{tabular}
    \label{tab:full-table2}
\end{table}

\begin{table}
    \centering
    \renewcommand\thetable{A1 (cont)}
    \caption{Table A1 continued. The extinction in filter $X$, $A_{X}$, assumes $A_{V} = A_{V, griz}$ and $R_{V} = 2.61$.}
    \begin{tabular}{c|c|c|c|c|c|c|c|c|c|c|c|c}
        SNID & \multicolumn{4}{c}{Peak App. Magnitude [mag]} & \multicolumn{4}{c}{Peak App. Magnitude [mag]} & \multicolumn{4}{c}{Dust Extinction} \\
         & \multicolumn{4}{c}{(Dust Corrected)} & \multicolumn{4}{c}{(Dust Corrected, Stretch Corrected)} & \multicolumn{4}{c}{} \\
         & $g$ & $r$ & $i$ & $z$ & $g$ & $r$ & $i$ & $z$ & $g$ & $r$ & $i$ & $z$ \\
        \hline
        \hline
        2016W & $15.701 \pm 0.009$ & $15.739 \pm 0.019$ & $16.228 \pm 0.011$ & $16.343 \pm 0.008$ & $15.422 \pm 0.066$ & $15.582 \pm 0.032$ & $16.139 \pm 0.019$ & $16.181 \pm 0.042$ & 0.271 & 0.179 & 0.129 & 0.104 \\
        2016afk & $17.625 \pm 0.032$ & $17.663 \pm 0.035$ & $18.282 \pm 0.019$ & $18.370 \pm 0.047$ & $17.537 \pm 0.063$ & $17.644 \pm 0.025$ & $18.335 \pm 0.024$ & $18.394 \pm 0.057$ & 0.287 & 0.189 & 0.137 & 0.110 \\
        2020adkc & $18.334 \pm 0.199$ & $18.583 \pm 0.028$ & $19.286 \pm 0.084$ & $19.275 \pm 0.118$ & $18.283 \pm 0.402$ & $18.723 \pm 0.102$ & $19.298 \pm 0.113$ & $19.463 \pm 0.142$ & 0.140 & 0.092 & 0.067 & 0.054 \\
        2020evu & $17.711 \pm 0.021$ & $17.799 \pm 0.083$ & $18.602 \pm 0.100$ & $18.420 \pm 0.047$ & $17.826 \pm 0.030$ & $17.835 \pm 0.123$ & $18.688 \pm 0.112$ & $18.504 \pm 0.029$ & 0.548 & 0.362 & 0.261 & 0.210 \\
        2020ewx & $17.857 \pm 0.098$ & $17.932 \pm 0.029$ & $18.624 \pm 0.049$ & $18.618 \pm 0.044$ & $18.001 \pm 0.188$ & $18.096 \pm 0.089$ & $18.674 \pm 0.061$ & $18.750 \pm 0.078$ & 0.174 & 0.115 & 0.083 & 0.067 \\
        2020fhs & $15.563 \pm 0.087$ & $15.748 \pm 0.019$ & $16.152 \pm 0.059$ & $16.386 \pm 0.013$ & $15.190 \pm 0.163$ & $15.461 \pm 0.029$ & $15.983 \pm 0.071$ & $16.147 \pm 0.046$ & 0.621 & 0.410 & 0.296 & 0.238 \\
        2020ilb & $17.369 \pm 0.052$ & $17.428 \pm 0.050$ & $17.982 \pm 0.077$ & $18.019 \pm 0.065$ & $17.496 \pm 0.119$ & $17.525 \pm 0.166$ & $17.979 \pm 0.111$ & $17.929 \pm 0.181$ & 0.678 & 0.447 & 0.323 & 0.260 \\
        2020ioz & $18.048 \pm 0.069$ & $18.222 \pm 0.030$ & $18.851 \pm 0.055$ & $18.837 \pm 0.046$ & $18.173 \pm 0.176$ & $18.366 \pm 0.051$ & $18.870 \pm 0.056$ & $19.067 \pm 0.074$ & 0.081 & 0.054 & 0.039 & 0.031 \\
        \hline
    \end{tabular}
    \label{tab:full-table3}
\end{table}

\end{landscape}